\newif\ifacmconference
\acmconferencefalse

\newif\ifsubmission
\submissiontrue

\documentclass[lettersize,journal]{IEEEtran}

\AtBeginDocument{%
  \providecommand\BibTeX{{%
    \normalfont B\kern-0.5em{\scshape i\kern-0.25em b}\kern-0.8em\TeX}}}

\ifacmconference
    

\else

\fi

\usepackage{xpatch}
\usepackage{booktabs, multirow, makecell}
\usepackage{graphicx}
\usepackage{subcaption}
\usepackage[inline]{enumitem}
\usepackage{xcolor, soul}
\usepackage{tikz}
\usepackage{array}
\usepackage{tagging}
\usepackage{MnSymbol}
\usepackage{makecell}
\usepackage{url}
\usepackage{hyperref}
\usepackage{orcidlink}
\usepackage{amsmath}
\usepackage{algorithm}
\usepackage{algpseudocode}

\usepackage{environ}

\NewEnviron{review}{%
  \ifsubmission
    \textcolor{black}{\BODY}%
  \else
    \textcolor{blue}{\BODY}%
  \fi
\ }

\usepackage{cleveref}
\crefname{section}{Sec.}{Secs.}
\crefname{figure}{Fig.}{Figs.}
\crefname{table}{Table}{Tables}
\crefname{algorithm}{Algorithm.}{Algorithms.}

\usepackage{xspace}
\newcommand{\sysname}{\textsc{ChamaleoNet}\xspace}

\newcommand{\pktrx}{\textsf{\small pkt\_rx}\xspace}

\usepackage{pifont}
\usepackage{xcolor}

\usepackage[acronym]{glossaries}
\glsdisablehyper
\newacronym{nf}{FSD-NF}{Flow State Detection Network Function}
\newacronym{dt}{DT}{D\emph{etection} T\emph{imeout}}
\newacronym{sdn}{SDN}{Software-Defined Networking}
\newacronym[longplural={Intrusion Detection/Prevention Systems}]{idps}{IDPS}{Intrusion Detection/Prevention System}
\newacronym{cdn}{CDN}{Content Delivery Network}
\newacronym{ddos}{DDoS}{Distributed Denial-of-Service}
\newacronym{ebpf}{eBPF}{extended Berkeley Packet Filter}
\newacronym{xdp}{XDP}{eXpress Data Path}
\newacronym{mat}{MAT}{Match-Action Table}
\newacronym{bri}{BRI}{Barefoot Runtime Interface}
\newacronym{dpo}{DPO}{Data Protection Officer}
\newacronym{nat}{NAT}{Network Address Translation}
\newacronym{fnr}{FNR}{False Negative Rate}

\ifsubmission
    \newcommand{\paolo}[1]{}
    \newcommand{\andrea}[1]{}
    \newcommand{\ac}[1]{}
    \newcommand{\mm}[1]{}
    \newcommand{\id}[1]{}
    \newcommand{\zw}[1]{}
\else
    \newcommand{\paolo}[1]{\textit{\color{magenta}[PG: #1]}}
     \newcommand{\andrea}[1]{\textit{\color{olive}[AB: #1]}}
    \newcommand{\ac}[1]{\textit{\color{brown}[AC: #1]}}
    \newcommand{\mm}[1]{\textit{\color{purple}[MM: #1]}}
    \newcommand{\id}[1]{\textit{\color{red}[ID: #1]}}
    \newcommand{\zw}[1]{\textit{\color{orange}[ZW: #1]}}
\fi

\usepackage[font=small,format=plain,labelfont=bf,textfont=bf]{caption}

\newcommand{\smartparagraph}[1]{{\bf #1}.\ }

\newcommand{\circled}[1]{\tikz[baseline=(myanchor.base)] \node[circle,fill=.,inner sep=1pt] (myanchor) {\color{-.}\bfseries\footnotesize #1};}

\title{\sysname: Programmable Passive Probe\\ for Enhanced Visibility on Erroneous Traffic}

\begin{document}

\author{Zhihao Wang\orcidlink{0000-0003-1989-3067}, Alessandro Cornacchia\orcidlink{0000-0002-4734-3321}, Andrea Bianco\orcidlink{0000-0002-5903-3558},~\IEEEmembership{Senior Member,~IEEE}, Idilio Drago\orcidlink{0000-0003-1932-1261}, Paolo Giaccone\orcidlink{0000-0003-4283-7936},~\IEEEmembership{Senior Member,~IEEE}, Dingde Jiang\orcidlink{0000-0003-0284-5624},~\IEEEmembership{Member,~IEEE}, and Marco Mellia\orcidlink{0000-0003-1859-6693},~\IEEEmembership{Fellow,~IEEE}

\thanks{Zhihao Wang is with University of Electronic Science and Technology of China, 611731 Chengdu, China, and the Department of Control and Computer Engineering, Politecnico di Torino, 10129 Turin, Italy (e-mail: zhihao.wang@polito.it).

Alessandro Cornacchia is with the Computer, Electrical and Mathematical Sciences and Engineering Division, King Abdullah University of Science and Technology, 23955 Thuwal, Saudi Arabia (e-mail: alessandro.cornacchia@kaust.edu.sa).

Andrea Bianco and Paolo Giaccone are with the Department of Electronics and Telecommunications, Politecnico di Torino, 10129 Turin, Italy (e-mail: andrea.bianco@polito.it; paolo.giaccone@polito.it).

Idilio Drago is with the Computer Science Department, Università di Torino, 10124 Turin, Italy (e-mail: idilio.drago@unito.it).

Dingde Jiang is with the School of Information and Communication Engineering, University of Electronic Science and Technology of China, 611731 Chengdu, China (e-mail: jiangdd@uestc.edu.cn).

Marco Mellia is with the Department of Control and Computer Engineering, Politecnico di Torino, 10129 Turin, Italy (e-mail: marco.mellia@polito.it).
}
}

\markboth{Journal of \LaTeX\ Class Files,~Vol.~14, No.~8, August~2021}%
{Shell \MakeLowercase{\textit{et al.}}: A Sample Article Using IEEEtran.cls for IEEE Journals}


\maketitle

\begin{abstract}
Traffic visibility remains a key component for management and security operations. Observing erroneous traffic, i.e., \textit{unanswered requests or error messages}, is fundamental to detecting misconfiguration, temporary failures or attacks. \sysname transforms any production network into a transparent monitor to let administrators collect such erroneous traffic. \sysname is programmed to ignore well-formed traffic and record only erroneous packets, including those generated by misconfigured or infected internal hosts, and those sent by external actors that scan for services.
Engineering such a system poses several challenges, from scalability to privacy. Leveraging the \gls{sdn} paradigm, \sysname processes the humongous amount of traffic flowing through the network border and focuses on erroneous packets only, lowering the pressure on the collection system. Moreover, it offers traffic anonymisation to conform to privacy regulations. 
\sysname enables the seamless integration with active deceptive systems like honeypots that can impersonate hosts/ports/services and engage with senders.
In an operational scenario, we show that the SDN in-hardware filtering reduces the traffic to the controller by 90\%, resulting in a scalable solution, which we offer as open source. 
Simple statistical analytics unveil the precious information carried by erroneous traffic. We discover internal misconfigured and infected hosts, identify temporary failures, and show enhanced visibility on attackers' scanning activities that look for vulnerable services.
\end{abstract}


\section{Introduction}
\label{sec:intro}

Network monitoring has always been the first step to implement network management or security policy.
Some monitoring systems offer visibility on regular traffic to derive statistics on performance and application usage, e.g., flow loggers~\cite{trevisan2016dpdkstat,deri2014ndpi,zhang2019}; some systems focus on protecting the network infrastructure, e.g., \begin{review}\glspl{idps}\end{review}~\cite{abdulganiyu2023systematic}; some collect unsolicited traffic, e.g., network telescopes~\cite{pang2004characteristics,MooreNetworkTelescopesTechnical2004}. 

Yet live networks contain well-formed flows, i.e., requests that get answered by the destination, and flows that remain unanswered or generate error messages at the network layer. 
In this paper we focus on the latter, which we term \textit{erroneous traffic}.
Such traffic arises when clients attempt to connect to offline (or firewalled) systems, or try to reach unintended or unavailable servers (possibly caused by routing anomalies). At last, malicious actors scanning networks and services generate erroneous traffic.

Prior work~\cite{PauleyDScopeCloudNativeInternet2023,RichterScanningScannersSensing2019a,metatelescope} has shown erroneous traffic reveals interesting insights. 
For example, 
\cite{RichterScanningScannersSensing2019a} shows that unsolicited traffic in large \begin{review}\glspl{cdn}\end{review} is not uniformly random: over 30\% of scans are localized, targeting few address regions. Similarly, DScope~\cite{PauleyDScopeCloudNativeInternet2023} reports up to $450\times$ more scan traffic than expected under random scanning. These results indicate that erroneous traffic in production networks reflects systematic targeting and structural bias rather than background noise. 
This motivates detecting and collecting such traffic in production environments to study targeting effects and distribution patterns. 
Our previous work~\cite{sordello2025poster} also reveals the value of outbound erroneous traffic in identifying signals of misconfiguration, transient failures, or attacks.

\begin{review}
In this paper, we design a passive traffic collector explicitly targeted to erroneous traffic.
We identify three key requirements that an ideal collector should satisfy:
\begin{enumerate*}[label=\emph{R\arabic*}]
    \item \label{req:completeness}\emph{Completeness:} the collector should provide full coverage of erroneous traffic, even when error status becomes evident only at runtime;
    \item \label{req:live}\emph{Transparency:} the collector must operate on live address space, coexisting with legitimate address assignments and active services without requiring dedicated dark subnets;
    \item \label{req:privacy}\emph{Data minimisation:} regulatory and institutional policies~\cite{gdpr,ccpa} require that capture focuses on traffic relevant to security or operations, avoiding bulk logging of legitimate sessions.
\end{enumerate*}
We argue that none of the existing monitors jointly satisfies the three requirements.
\end{review}
Traditional network telescopes record all packets destined to an \textit{unused subnet}, i.e., where no production host is connected, and thus are inherently designed to capture erroneous traffic.
\begin{review}
However, they present two limitations.
\end{review}
First, because telescopes operate on reserved and unassigned address spaces, devoting precious IP subnets solely for data collection, they cannot operate on live address space, 
\begin{review}violating \ref{req:live}.\end{review}
Flexible telescopes~\cite{PauleyDScopeCloudNativeInternet2023,RichterScanningScannersSensing2019a,metatelescope} exist, but still require addresses to remain unassigned during collection.
Second, telescopes only capture traffic directed to the monitored dark space (e.g., inbound Internet radiation~\cite{pang2004characteristics}), missing erroneous traffic directed to live hosts and all outbound erroneous traffic generated by internal hosts, 
\begin{review}
    thus violating \ref{req:completeness}.    
\end{review}
Traditional flow loggers operate on live networks but provide only statistical summaries of flows, without distinguishing erroneous traffic. 
\begin{review}
Therefore, they violate \ref{req:completeness} (e.g., due to sampling) and \ref{req:privacy}.
\end{review}
Finally, \acrfullpl*{idps} inspect packets against predefined rules or known signatures, \begin{review}falling short of \ref{req:completeness} when erroneous traffic is not known a priori, and of \ref{req:privacy} when bulk logging is enabled to capture unknown traffic.\end{review} 
Although not designed to collect erroneous traffic, \glspl{idps} can, in principle, be extended to track it by maintaining per-flow states (e.g., triggering events for new requests and timeouts~\cite{ergenc2023tsnzeek}), since they inherently support stateful processing. However, stateful operation is costly, and their support for raw packet logging is limited: the scale of typical data rates challenges the usage of \gls{idps} to log erroneous traffic~\cite{trevisan2016dpdkstat,zhang2019,zeek2025newpacket}.


To close this gap, we propose \sysname, a flexible monitoring system that transforms any production network into a 
programmable probe for erroneous traffic.
\begin{review}
    Unlike traditional network telescopes, \sysname avoids reserving IP addresses for monitoring purposes and dynamically detects erroneous traffic exchanged by services in an operational, live network, and opportunistically collects it.
\end{review} 

\begin{figure}[t]
    \centering
    \includegraphics[scale=0.7]{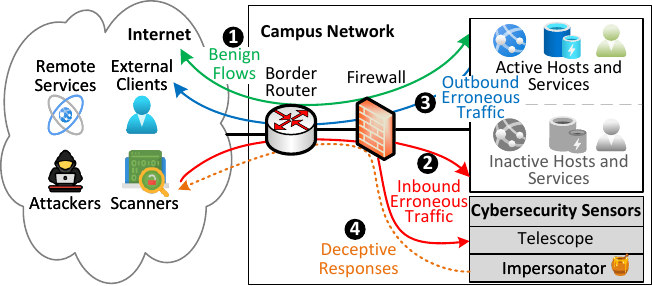}
    \caption{\begin{review}Concept of erroneous traffic and scope of \sysname in a campus network, highlighting the actors and traffic types.\end{review}
    }
    \label{fig:background}
\end{figure}

\begin{review}
\cref{fig:background} depicts the concept of erroneous traffic and the scope of \sysname.
\end{review}
\sysname ignores all regular traffic (green arrow \circled{1}), and only collects those packets for which no valid response is observed (arrows \circled{2} \circled{3}). 
More specifically, \sysname logs the erroneous traffic generated by \textit{external hosts} (inbound erroneous traffic \circled{2}), offering visibility on Internet radiation and attackers' scanning activities, and enabling the administrators to engage with scanners by selectively enabling impersonators.
Likewise, \sysname logs requests originated by  \textit{internal hosts} (outbound erroneous traffic \circled{3}), unveiling issues with external services and offering visibility on misconfigured or infected internal hosts, thereby facilitating troubleshooting and repair. 
At last, \sysname can transform itself into an active impersonator (orange dashed arrow \circled{4}), impersonating inactive hosts or services, as a honeypot system.

Engineering \sysname requires ingenuity. We leverage \acrfull*{sdn} and programmable switches to transform a live network into a flexible monitor.
To identify erroneous traffic (\circled{2} and \circled{3}), \sysname tracks unanswered requests or error messages destined to addresses assigned to unreachable internal hosts.
Once identified, \sysname steers erroneous traffic to back-end cybersecurity sensors, whether they are simple passive collectors or active impersonators that can impersonate the destination. 
The \gls{sdn} architecture naturally maps to these design requirements. The controller handles the stateful detection logic that identifies non-responding services. Meanwhile, programmable switches perform line-rate filtering of benign traffic, shielding the controller from the full traffic volume.
At the same time, network programmability enables flexible steering of erroneous traffic towards cybersecurity sensors, and offloads privacy-aware packet processing functionalities (e.g., packet anonymisers) to address regulatory requirements to the network data plane. 

In summary, 
\sysname transforms any network into a flexible and transparent monitor for erroneous traffic, enabling visibility on a wide range of management and cybersecurity events coming from external and internal hosts. 
We deployed  \sysname at our University campus network for over eight months, and report on the key findings.
We offer \sysname to the community as open source.\footnote{\url{https://github.com/zhihao1998/ChamaleoNet}}

Our contributions are as follows:
(i) We introduce the concept of erroneous traffic and design \sysname, a system that transforms any private network into a flexible monitor of erroneous traffic.
(ii) We implement \sysname atop SDN principles to transparently filter, anonymise, and steer traffic, respecting users' privacy. Offloading the filtering of regular traffic to the switch reduces the controller load by 90\%. 
(iii) \sysname improves visibility on external Internet radiation while empowering the administrators to engage with scanners by selectively enabling impersonators.
(iv) Simple statistical analysis of the erroneous traffic patterns expose suspicious internal hosts, misconfigured systems, and routing issues, which would otherwise go unnoticed.

Next, \cref{sec:related} presents the related work. \cref{sec:arch} presents the design and architecture of \sysname. \cref{sec:impl} describes its implementation. \cref{sec:action} reports on its deployment and key findings, while \cref{sec:perf} evaluates its performance in a controlled environment. \cref{sec:discussion} outlines strategies to improve the architecture. We discuss the ethical considerations in \cref{sec:ethical} and conclude the paper in \cref{sec:conclusion}.

\section{Related Work}\label{sec:related}

\subsection{Passive Measurement Mechanisms}\label{sec:related:passive-measurement}

Several mechanisms exist to passively monitor traffic, from simple network telescopes to IDS/IPS or IDPS. Telescopes collect packets destined to dark subnets with no active hosts. The collected Internet background radiation~\cite{pang2004characteristics,MooreNetworkTelescopesTechnical2004} gives visibility on Internet scanners, botnets, coordinated attacks, failures, routing issues and even censorship policies~\cite{jonker_millions_2017,antonakakis_understanding_2017}. Telescopes ignore regular and some erroneous traffic and waste resources due to the need to leave unused large ranges of IP addresses~\cite{caida2024}.

Recent works have explored flexible telescope designs. 
In~\cite{RichterScanningScannersSensing2019a} authors leveraged CDN infrastructure to study unsolicited traffic reaching production servers, collecting packets blocked at CDN firewalls. Attracted by live CDN nodes, attackers send a variety of packets that are not observed in telescopes. 
Similarly, DScope~\cite{PauleyDScopeCloudNativeInternet2023} places telescopes in cloud data centers. It continually and dynamically allocates new IP addresses from the Amazon Web Services, retaining them for 10 minutes to collect inbound TCP connection attempts. Meta-telescope~\cite{metatelescope} infers unused Internet address blocks and captures the corresponding traffic without requiring administrative control over those blocks. 

These designs require that addresses remain \textit{unassigned/unused} when used for collecting unsolicited traffic. As such, they lack visibility into: \textit{i)} addresses that are temporarily offline but not yet incorporated into the pool, and \textit{ii)} addresses with only a subset of ports active. In contrast, \sysname immediately gains visibility of a host once it goes dark and continuously monitors the unused ports of \textit{active} hosts.

Reactive telescopes such as Spoki~\cite{HiesgenSpokiUnveilingNew2022} and others~\cite{SoroEnlighteningDarknetsAugmenting2023} augment telescopes with the ability to respond to incoming requests through simple impersonators, e.g., opening the TCP connection to capture the first payload.

On the other extreme, IDPSes, e.g., Snort, Suricata and Zeek, analyse live traffic in real time and block malicious or suspicious activities, eventually alerting administrators~\cite{abdulganiyu2023systematic}. 
IDPSes react to attacks and system abuses, using known signatures or anomaly detection. They offer the ability to collect packet traces, selectively logging only packets causing the incident~\cite{HADEM2021108015} or logging all packets.

In between, flow loggers offer visibility on operational traffic~\cite{trevisan2016dpdkstat,deri2014ndpi,zhang2019}, processing packets in real-time to export flow-level statistics, but provide only limited packet capture capabilities, typically for traffic samples or specific protocols.

\begin{table}[!tb]
\centering
\caption{Scope of \sysname}
\label{tab:scope}
\resizebox{\columnwidth}{!}{%
\begin{tabular}{l|c|c|c|c|c|c|c|c|}
\cline{2-8}
 &  {\textbf{Traffic}} & {\textbf{Focus}} & {\textbf{Pkt}} & {\textbf{Flow}} & {\textbf{Alert}} & {\textbf{Reply}}& {\textbf{Rules}} \\ \hline
\multicolumn{1}{|l|}{\textbf{Telescope}} & In  & Dark & $\checkmark$ &  &  &  & static \\ 
\hline
\multicolumn{1}{|l|}{\textbf{\begin{tabular}[c]{@{}l@{}}Reactive\\ Telescope\end{tabular}}} & In  & Dark & $\checkmark$ &  &  & $\checkmark$ & static \\ 
\hline
\multicolumn{1}{|l|}{\textbf{\begin{tabular}[c]{@{}l@{}}Flow\\ monitor\end{tabular}}} & In/Out & All &  & $\checkmark$ &  & & static\\ 
\hline
\multicolumn{1}{|l|}{\textbf{IDPS}} & In/Out & Malicious & $\checkmark$ & $\checkmark$ & $\checkmark$ & & static \\ 
\hline
\multicolumn{1}{|l|}{\textbf{\sysname}} & In/Out & Erroneous & $\checkmark$ &  &  & $\checkmark$ & dynamic \\ 
\hline
\end{tabular}
}
\end{table}

~\Cref{tab:scope} summarises the \sysname scope. All platforms process incoming traffic. Flow monitors, IDPSes and \sysname offer visibility on traffic initiated by internal hosts (outgoing traffic). Telescopes focus on dark traffic, flow monitors summarize all traffic, while IDPSes are usually set to look for malicious traffic. Telescopes capture packets, while flow monitors log information at the flow level. IDPSes produce alerts based on rules, and have some packet capture abilities. Only reactive telescopes and \sysname support backend impersonators that can actively respond to incoming requests, such as completing the TCP handshake.

\subsection{Monitoring and Defence Mechanisms in SDN}\label{sec:related:sdn-monitor}

The authors of~\cite{survey24} provide a comprehensive taxonomy of past work that leveraged the use of SDN as a defence mechanism. Closest to our work are those related to honeypot systems. The study in~\cite{li219} introduced a honeynet based on SDN, akin to \sysname, which duplicates traffic to the honey-network for further examination. In~\cite{kara195}, authors proposed a solution where the controller inspects suspicious traffic packets and redirects it to honeypots. 
None of these works can dynamically filter well-formed traffic at runtime and redirect only \emph{erroneous} traffic to collector systems.

\subsection{Stateful Traffic Processing with Programmable Switches}

\sysname enables stateful processing of incoming traffic by leveraging the SDN switch and virtualized network functions. Some past works have focused on implementing stateful processing only in the switch, enabled by \begin{review}Programming Protocol-independent Packet Processors (P4)\end{review}~\cite{p4paper}. 
\cite{icc17} proposed a reactive traffic control application that redirects traffic in real-time to a traffic classification engine, \begin{review} entirely within the SDN switch. \cite{pao19} leverages a stateful approach directly in data plane to detect TCP SYN flood attacks. \cite{sdn17} uses SDN switch to mirror traffic to IDS engines, where the controller may instruct the switch to block malicious flows. Unlike \sysname, it neither buffers packets while the decision is being made, including the critical first packet, nor considers erroneous or internally generated traffic.
\end{review}


\smartparagraph{Summary} 
\sysname is the only system that targets erroneous traffic. It places itself between passive telescopes and advanced flow monitors and IDPSes, offering visibility on both external and internal erroneous traffic.
\sysname differs from previous works in several key aspects:
\begin{enumerate*}[noitemsep, leftmargin=*]
    \item  focuses on erroneous traffic, which previous systems largely ignore;
    \item  leverages SDN programmability to detect erroneous traffic through dynamic rules;
    \item  provides visibility into both incoming and outgoing traffic;
    \item  operates on live networks and transparently treats unused addresses -- including temporarily offline hosts -- as dark space;
    \item  extends the scope of dark space to the service level, i.e., unused ports on otherwise active servers;
    \item  logs traffic at the packet level while preserving privacy;
    \item  can selectively steer traffic to impersonators which can impersonate inactive hosts or specific services.
\end{enumerate*}


\section{System Design}
\label{sec:arch}

\cref{fig:controller_arch} shows an overview of \sysname. We design it as a pluggable network system (blue box) with minimal configuration. Connected to the campus network border router(s), \sysname observes all traffic entering/leaving the network and forwards only the erroneous portion to the collectors, optionally letting collectors impersonate the destination host or service.
Upon observing the first packet of a new flow\footnote{We define a flow via the usual 5-tuple, up to the transport layer. Flows are bi-directional: packets matching the same 5-tuple belong to the same flow, regardless of their direction.}, 
it must determine whether the destination service is active, and eventually forward traffic destined to inactive services or hosts to the backend. \sysname must not cause any disruption or interference to the active services, nor forward production traffic payload to the collectors. Hence, we design the system to be \emph{transparent} to services and benign traffic.

\subsection{\sysname Architecture}
\label{sec:design}

\sysname leverages the SDN paradigm to separate the data plane, responsible for line-rate packet filtering, from the control plane, which manages the stateful logic to distinguish erroneous from benign traffic. 

We here assume an out-of-path deployment (see~\cref{sec:discussion} for a discussion of the in-path alternative). All ingress and egress traffic, destined to and coming from the campus network, is mirrored to an SDN switch via either span ports or physical layer splitters. In case of multiple upstream providers, we assume all traffic is forwarded to handle asymmetric routing. 

\begin{figure}[t]
    \centering
    \includegraphics[width=\columnwidth]{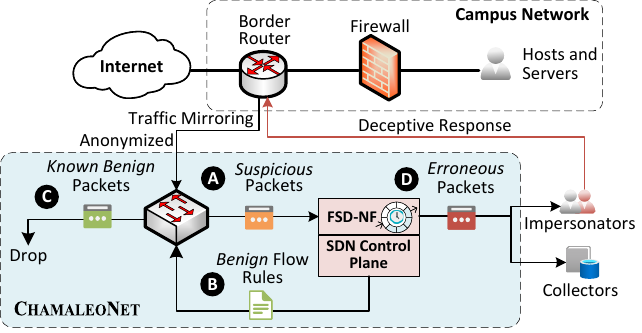}
    \caption{Overview of \sysname architecture.}
    \label{fig:controller_arch}
\end{figure}

\begin{review}
    The switch applies dynamic per-service filters.
\end{review}
When a packet arrives and does not match any filter, the switch forwards it to the SDN controller. Such a packet may either belong to a new legitimate flow or be potentially erroneous. Since both cases are plausible a priori, initially we denote these packets as \emph{suspicious packets} (event \circled{A} in \cref{fig:controller_arch}).

\sysname is designed to minimize logging of traffic destined to active services. To achieve this, a dedicated {\em \acrfull*{nf}} determines whether a suspicious packet is \emph{answered} (the service is active) or \emph{unanswered} (the service is temporarily inactive). 
\acrshort*{nf} buffers the suspicious packet for a pre-defined DeTection (DT) timeout during which it waits for a response packet of the same flow. 
Two cases can occur:
\begin{enumerate}[noitemsep, topsep=0pt,leftmargin=*]
\item A response packet is received within the \acrshort*{dt} timeout: the service is deemed \textit{active} and the suspicious packet (and following packets) are part of a legitimate flow.
The controller drops the packet and installs a flow rule on the SDN switch (\circled{B}) to drop all matching packets. The SDN switch will discard any subsequent packet of the benign flow, avoiding overflowing the controller (\circled{C}).
\item The DT timeout expires: the service is deemed \textit{inactive} and the suspicious packet is classified as \emph{erroneous}. The controller forwards the original packet buffered at the \acrshort*{nf} to the cybersecurity collectors (\circled{D}), which may log or eventually reply to such packets acting as a honeypot.
\end{enumerate}

Albeit easy to implement, \sysname treats ICMP unreachable errors as erroneous rather than valid responses.

By observing all initial flow packets and responses, \sysname can keep track of which internal hosts are \textit{alive}: Whenever the sender IP address corresponds to an internal host, \sysname marks the sending host as alive. 
This allows it to distinguish cases where external requests go unanswered because the destination is not alive (or dark, as in the case for telescopes) or present but refusing to answer (e.g., firewalled service or rebooting hosts). \sysname can take decisions on whether to respond to a packet, log or ignore it accordingly. 

\subsection{Running Example}

\begin{figure}[t]
    \centering
     \includegraphics[width=0.7\columnwidth]{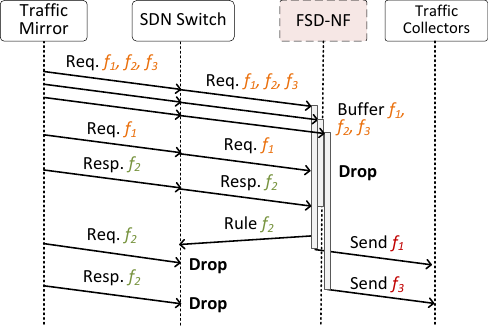}
    \caption{\sysname detecting two flows with erroneous packets ($f_1$, $f_3$) and one benign flow ($f_2$).}
    \label{fig:host-flow-diagram}
\end{figure}

The time diagram in \cref{fig:host-flow-diagram} shows an example of how \sysname works.
The switch receives traffic from the traffic mirror. It receives three packets belonging to three different flows ($f_1, f_2, f_3$).
No rule matches these packets, so the switch forwards them to the FSD-NF, which buffers them and sets a DT timer to wait for more packets of the same flow to eventually arrive.
Next, the FSD-NF observes a second request packet belonging to $f_1$ (e.g., a retransmission from the same sender), and drops it. Eventually, this second packet could be stored at the FSD-NF and sent to the cybersecurity collector too. Then, a response packet for $f_2$ arrives and is forwarded to FSD-NF, which deems the service and the host as active. FSD-NF marks the flow $f_2$ as ``benign'' and installs a rule in the switch to drop all subsequent packets belonging to $f_2$. In the meantime, FSD-NF drops all $f_2$ packets that may still be forwarded by the switch. 
At last, $f_1$ and $f_3$ DTs expire. The \acrshort*{nf} sets the corresponding buffered packets as ``erroneous'' and sends them to the collectors, removing any internal state related to them.

\begin{figure*}[!t]
\centering
\begin{tikzpicture}
    \node[anchor=south west, inner sep=0] (img) at (0,0)
      {\includegraphics[width=0.85\textwidth]{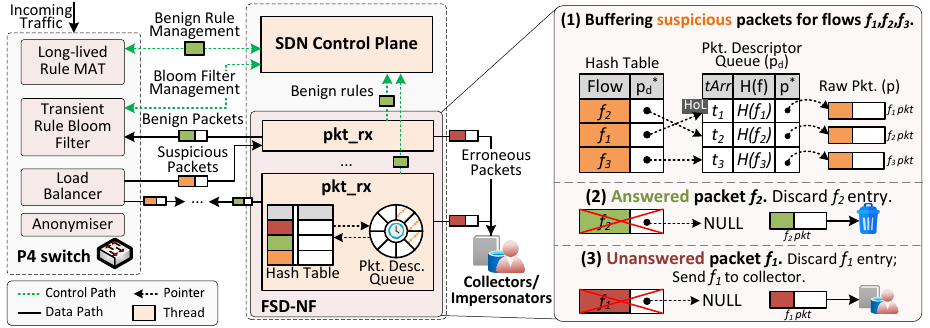}};

    \begin{scope}[x={(img.south east)}, y={(img.north west)}]
        \node at (-0.026, 0.8)
          {\textcolor{black}{\scriptsize{\S \ref{sec:mat_filtering}}}};
        \node at (-0.026, 0.61)
          {\textcolor{black}{\scriptsize{\S \ref{sec:bloom_filter}}}};
        \node at (-0.026, 0.31)
          {\textcolor{black}{\scriptsize{\S \ref{sec:anonymizer}}}};
        \node at (0.57, 0.85)
          {\textcolor{black}{\scriptsize{\S \ref{sec:nf}}}};
    \end{scope}
\end{tikzpicture}
\caption{\begin{review}Implementation of \sysname and its user-space packet processing workflow.\end{review}}
\label{fig:implementation}
\end{figure*}

\subsection{Assumptions and Threat Model}\label{sec:design:assumptions}

\smartparagraph{Assumptions}
We make the following assumptions: 
\begin{enumerate*}[noitemsep, leftmargin=*]
    \item \emph{Unified visibility:} the monitored network has unified borders and all ingress/egress traffic is mirrored to \sysname (out-of-path); 
    \item \emph{Asymmetric routing:} in multi-homing settings, traffic is mirrored/forwarded such that asymmetric return paths remain visible. 
    Note that asymmetric routing is an open challenge for such monitoring systems, e.g., in Meta-TeleScope~\cite{metatelescope};
    \item \emph{Timing consistency:} packet timestamps and ordering at the monitoring vantage point are sufficiently accurate for enforcing the observation window;
    \item \emph{No inbound spoofing:} border devices are trusted and enables common filtering that prevents external hosts from spoofing internal source IPs (e.g., BCP38~\cite{bcp38}).
    \item \emph{Dynamic SDN-like rule support in the data plane:} the switch component---whether a physical SDN/P4 switch, a SmartNIC, or an \begin{review}\gls{ebpf}/\gls{xdp}\end{review} program---accepts dynamic flow rules and forwards unmatched packets to the controller.
\end{enumerate*}

\smartparagraph{Threat model}
\begin{review}
\sysname captures interactions in which requests reach routable addresses, but no observable reverse-direction response is detected within a defined observation window.
\end{review}
Such interactions arise when endpoints are inactive, filtered, unreachable at the service layer, misconfigured, or non-responsive.
The threat model, therefore, includes network activities interacting with non-responsive endpoints. Examples include scanning, probing, exploitation attempts, and unintended communications. These activities may be malicious or benign in intent. The defining property is the absence of an observable response within the observation window, not the intent of the sender. The goal of \sysname is to identify and analyse such interactions without disrupting normal traffic forwarding.

%
\begin{review}
    \smartparagraph{Monitoring blind spots and evasion}
\end{review}
The threat model is limited to L3/L4 response observability within a configurable observation window.
\begin{review}
Interactions that successfully establish bidirectional communication within the observation window fall outside the erroneous-traffic model.
An attacker could evade detection by injecting crafted reverse-direction packets that traverse the monitored border within the DT timeout.
\end{review}
This requires coordinated control over the reverse path (e.g., control of both endpoints or reverse-path injection capability).
\begin{review}
In such cases, the resulting interaction becomes indistinguishable from legitimate L3/L4 communication under \sysname’s observability model. Detecting such behaviours is outside the scope of \sysname and is typically delegated to other monitoring mechanisms, such as IDPSes.
\end{review}


\section{Implementation}
\label{sec:impl}

\cref{fig:implementation} depicts the main components of \sysname and their interactions.
We first describe the \acrshort*{nf} packet processing logic and data structures (\cref{sec:nf}), then present a two-tier rule management strategy for filtering benign flows at the SDN switch (\cref{sec:impl:rule-management}), and finally highlight the traffic anonymizer (\cref{sec:anonymizer}).



\subsection{\texorpdfstring{\acrfull*{nf}}{Flow State Detection Network Function (FSD-NF)}}
\label{sec:nf}

We rely on state-of-the-art per-flow monitoring based on multi-threading to implement the \acrshort*{nf}, which has been shown to cope with several tens of Gbps on off-the-shelf hardware~\cite{trevisan2016dpdkstat}.
For packet reception, \sysname uses any library that allows to capture packets. In our prototype, we use \texttt{libpcap} configured with ``immediate mode'', which disables packet batching and delivers packets to the \acrshort*{nf} upon arrival. Any other kernel-bypass packet I/O frameworks, such as DPDK~\cite{dpdk}, are transparent to \sysname design, and orthogonal to the scope of this work.
At the \acrshort*{nf}, multiple \pktrx threads process suspicious packets arriving from the switch, buffering packets and executing the \acrshort*{nf} packet processing logic described next.
Suspicious packets are load-balanced across \pktrx threads via hash-based selection on internal service identifiers. This preserves per-service affinity because the hashing operation is offloaded to the switch.

\subsubsection{Packet buffering and data structures}
\cref{fig:implementation} shows the \sysname data structures and processing. 
There are three main data structures in FSD-NF: a hash table to store flow states; a packet descriptor queue to manage timeouts; a packet buffer to cache raw packets.  
The \acrshort*{nf} buffers suspicious packets and waits for a response (benign flow) or for the DT timer to expire (erroneous packet).
We store packets in FSD-NF memory and track their locations using descriptors organized as a ring buffer queue.
The descriptor stores also the packet arrival time {\em tArr} and a pointer to the entry in the flow hash table. 
Event (1) in~\cref{fig:implementation} (right) illustrates the relationship between the data structures for packets of flow $f_1, f_2$ and $f_3$, while \cref{alg:dt-timer} shows the pseudocode.
When a new packet arrives and does not match any entry in the hash table, we store it in memory, create a new packet descriptor, set the $tArr$ field to the current time, and insert the packet descriptor in the packet descriptor queue. To optimise the lookup when matching responses, we also store in the hash table a pointer to the packet descriptor position within the ring buffer. 

\subsubsection{Erroneous packet detection and DT timer management}
\sysname detects erroneous packets by waiting for the DT timeout to expire. To avoid the overhead of managing multiple timers, one per packet, we implement a lazy timeout mechanism with a single timer. We exploit the simple observation that DT timeouts for suspicious packets expire in the same order in which the packets are received, and manage expiration via the following polling mechanism embedded within the \pktrx thread (ln.\ref{ln:periodic-dt-check}).
The \pktrx thread periodically (every $P_{DT}$) scans the packet descriptor queue starting from the entry with oldest $tArr$. For each entry, \pktrx compares the current wall-clock time with {\em tArr}. 
If the elapsed time is greater than the DT timeout,
it deletes the hash table entry, sends the packet to the collector and frees the corresponding packet descriptor. 
We stop at the first descriptor that has not exceeded the DT timeout.
Event (3) in~\cref{fig:implementation} (right) illustrates a  packet of flow $f_1$ marked erroneous and sent to the collector. 
To limit the computation cost of this linear search operation, we allow a maximum search depth equal to $d_{\max}$ positions.\footnote{While this check could be done in a separate thread, it would create race conditions on the ring buffer, introducing synchronisation overheads. Experimental results confirm that it is not needed (see \cref{sec:perf}).}

\subsubsection{Benign flow detection} 

If FSD-NF receives a response packet before the DT expiration, the flow is deemed \emph{benign}---i.e., internal service is active (\cref{sec:design}). \acrshort*{nf} 
deletes the corresponding suspicious packet from the packet descriptor queue, and installs an entry in the switch to filter all future packets of this benign flow. 
Response packets are detected in constant time using hash table lookups, by matching 
the response packet's flow identifier to the corresponding suspicious packet's flow entry.


\begin{algorithm}[tb]
\caption{\pktrx and DT Timer Management in \acrshort*{nf}}
\label{alg:dt-timer}
\begin{algorithmic}[1]
\Require Packet descriptor queue $Q$, hash table $H$, DT timeout, polling period $P_{DT}$, max search depth $d_{\max}$
\Ensure Suspicious packets $p$ belonging to flow $f$ forwarded to collector, expired descriptors freed

\Statex \textbf{On new suspicious packet arrival $p$:}
\If{$p$ matches an existing entry in $H$}
    \If{$H_f.tResp \neq \texttt{null}$} \Comment{install pending}
        \State Drop $p$ and \Return
    \ElsIf{$p.dir = H_f.dir$} \Comment{same-direction duplicate}
        \State Drop $p$ and \Return
    \Else \Comment{reverse-direction (response) packet}
        \State Install a benign rule to the switch
        \State $p_d^* \gets \texttt{null}$
        \State Delete $H_f$ from hash table $H$
    \EndIf
\Else \Comment{suspicious first packet}
    \State Create packet descriptor $p_d$
    \State $p_d.tArr \gets \texttt{currentTime()}$
    \State Buffer packet $p$
    \State $p_d.p^* \gets addr(p)$ \Comment{pointer to buffered packet}
    \State Insert flow entry $H_f$ in $H$
    \State $H_f.p_d^* \gets addr(p_d)$ \Comment{pointer to packet descriptor}
    \State $p_d.H_f \gets H_f$ \Comment{key in hash table entry}
    \State Enqueue $p_d$ at tail of $Q$
\EndIf

\Statex \textbf{Periodic timeout check (every $P_{DT}$):} \label{ln:periodic-dt-check}
\State $depth \gets 0$
\State $p_d^* \gets$ head of $Q$ \Comment{oldest descriptor}
\State $now \gets \texttt{currentTime()}$
\While{$p_d^* \neq \texttt{null}$ \textbf{and} $depth < d_{\max}$}
    \If{$now - p_d^*.tArr > \text{DT timeout}$}
        \State Delete entry $H_f$ from hash table $H$
        \State Forward buffered packet $p$ to collector
        \State Dequeue and free $p_d^*$
        \State $p_d^* \gets$ next descriptor in $Q$
        \State $depth \gets depth + 1$
    \Else
        \State \textbf{break} \Comment{subsequent entries not yet expired}
    \EndIf
\EndWhile
\end{algorithmic}
\end{algorithm}

\subsection{Filtering Rules Management}
\label{sec:impl:rule-management}

The \acrshort*{nf} dynamically installs filtering rules in the switch so that subsequent packets belonging to benign flows are dropped.
Filtering rules take the form of exact-match rules in \begin{review}\glspl{mat}\end{review} or approximate matching in Bloom filters (BF). We refer to the mechanism as a dual-channel rule management.
A \emph{fast-path} channel quickly pushes filtering rules to the switch using in-band packets that update BFs via the switch dataplane. 
A \emph{slow-path} channel manages the installation of long-lived rules for well-known flows via the control plane.
The terminology ``fast'' and ``slow'' refers to the latency of rule installation in the switch, which is negligible for BFs and can be up to hundreds of milliseconds for MATs. 

We describe how we balance the use of these two channels to filter benign traffic while keeping the switch rule set manageable.

\subsubsection{Slow path: filtering long-lived benign flows with MAT rules}
\label{sec:mat_filtering}
In production networks, long-lived internal services (e.g., web services) always respond to inbound packets and consistently generate well-formed bidirectional traffic, i.e., benign flows. 
\sysname identifies long-lived benign services and installs persistent filtering rules on the P4 switch using MATs.
Because of stability and predictability, flows belonging to long-lived services are ideal candidates for the slow-path channel: 
since MAT memory has limited capacity and longer installation latency, we reserve MAT rules for filtering stable and well-known services. 

The SDN control plane manages both rule installation and removal. When \pktrx threads identify flows belonging to these services as benign, they trigger rule installation via the SDN control plane, which installs exact-match MAT rules keyed by \texttt{\small (internal IP, internal port, protocol)}. 
Due to limited MAT memory capacity, the control plane also implements rule eviction to delete idle entries for (likely) terminated flows.
We implement the entry eviction based on the idle notification mechanism provided by Intel Tofino chipset. 
Each MAT entry maintains a \texttt{TTL} field that decrements when no packet matches within a \texttt{\small Query\_interval}; any matching packet resets the \texttt{\small TTL}.
When the \texttt{\small TTL} reaches zero, the switch triggers an idle timeout notification to the controller, which also resets the corresponding host's aliveness status.
We employ batching both for installation and removal to reduce overhead. 

\subsubsection{Fast path: filtering transient benign flows with Bloom filters}
\label{sec:bloom_filter}
The fast-path is designed to cope with periods of high benign flow churn, where many new flows are initiated within short time intervals. In such scenarios, the switch would need to install numerous rules in a short time, which can outpace the speed of MAT rule installation over the slow-path channel and the MAT memory capacity.
A BF~\cite{spidermon} is a space-efficient probabilistic data structure that tracks membership in unbounded streams with constant-time query complexity, amenable for implementation on programmable switches.
In \sysname, BFs allow to (1) quickly install rules directly in the data plane without involving the control plane, and (2) filter the bulk of flows with a small memory footprint, albeit with some false positives whose impact can be made negligible (\Cref{sec:filtering_accuracy}). 


Standard BFs do not support deletion or stale entry removal, which is necessary for \sysname to remove stale benign flows and detect internal hosts that go dark. To address this, we adopt a rotating BF design that maintains $r$ identical BFs.
Insertions occur only in the currently active BF, while queries span all $r$ filters to determine whether a flow has appeared within the current time window or in any of the past $r-1$ time windows.
Every $T_{\text{bloom}}$, \sysname rotates the active BF and clears the \emph{oldest} one, thereby removing stale entries after at most $r \cdot T_{\text{bloom}}$. Consequently, BF benign rules have a maximum lifetime of $r \cdot T_{\text{bloom}}$.
After expiration, packets associated with those rules are treated as new suspicious packets and forwarded to the \acrshort*{nf} for re-evaluation.
By design, this overhead is negligible: transient flows typically complete before the rotation window expires, and persistent services are promoted to MAT rules via the slow path.

We favor rotation over alternatives, such as counting BFs \cite{hill2018tracking}, because the latter would require the controller to maintain per-flow state to track staleness and issue explicit deletions.


\subsubsection{Promotion of long-lived flows to MAT rules}\label{sec:bloomfilter:promotion}
we consolidate flow rules for long-lived benign flow into the slow path based on the flow lifetime. 
If a flow appears in $n_{\text{longlived}}$ consecutive BF's clearing events---i.e., the benign flow lifetime is at least $n_{\text{longlived}}\cdot r \cdot T_{\text{bloom}}$---the corresponding service is identified as long-lived and the control plane migrates its flow rule from the fast path to the slow-path by installing an exact-match MAT rule.
To track the number of consecutive BF rotations in which a flow $f$ appears, the control plane increases a counter every time it receives a new suspicious packet for the same flow, after the first fast-path rule for $f$ has been installed in the BF. If the counter reaches $n_{\text{longlived}}$, the control plane promotes the flow rule to the MAT and resets the counter.

\subsection{Traffic Anonymizer}
\label{sec:anonymizer}

\begin{review}
    Privacy regulations such as GDPR~\cite{gdpr} and CCPA~\cite{ccpa} treat IP addresses and application-layer payloads as potentially personal information. Since \sysname processes traffic generated by users connected to the campus network, it must access only the information required for its task.\footnote{Privacy policies have been validated by our \gls{dpo}.} To this end, \sysname implements configurable anonymisation directly in P4 for scalability, supporting internal IP address obfuscation and transport-layer payload removal.
\end{review}

We leverage the Tofino Native Architecture~\cite{tna,IntelTofinoSeries}: Packets first traverse the ingress pipeline before being buffered at the Traffic Manager and processed at the egress pipeline. 
\begin{review}
    Processing can occur at both stages, coordinated via metadata.
\end{review}

\begin{figure}[t]
    \centering
    \includegraphics[scale=0.9]{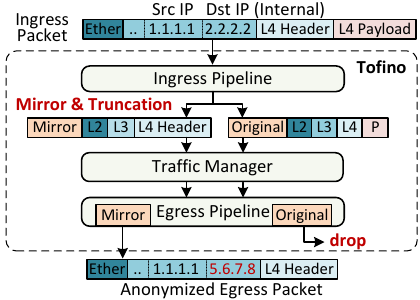}
    \caption{P4-based traffic anonymiser implemented on an Intel Tofino~\cite{IntelTofinoSeries} programmable chipset.}
    \label{fig:p4_anonymizer}
\end{figure}

\begin{review}
For IP address obfuscation of internal hosts, we rely on consistent obfuscation by XOR-ing the original value with a predetermined key salted with the last byte of the address itself. More complex techniques can be adopted~\cite{chen2020implementing}.

While limited payload processing is feasible on programmable ASICs~\cite{PayloadPark}, rewriting variable-length payload remains resource-constrained. Since \sysname aims to anonymise potentially sensitive L4 payload information, we implement L4-payload anonymisation through packet truncation using mirroring sessions, as shown in~\cref{fig:p4_anonymizer}. At ingress, the P4 program identifies packets requiring truncation, assigns them to a mirror session, and attaches internal metadata to distinguish the original packet from its mirrored copy, denoted as \texttt{\small Original} and \texttt{\small Mirror}, respectively. The controller configures the corresponding mirror session by specifying the truncation length applied at the Traffic Manager and the output port used for the mirrored packet. After truncation, both packets reach the egress pipeline, where the \texttt{\small Mirror} packet is forwarded to the output port connected to the \acrshort*{nf}, while the \texttt{\small Original} packet is dropped.
\end{review}


To truncate packets differently for different protocols, we define the truncation sizes either by matching the correct length in the protocol headers or by assigning pre-defined values (e.g., for unknown/unsupported protocols), each matching a different mirror session. 

\subsection{Exported metadata and live/dark host differentiation}
\label{sec:impl:metadata}

\begin{review}
Unlike traditional telescopes that monitor only reserved dark subnets, \sysname collects erroneous traffic directed to both alive and dark hosts. To support downstream security analyses (\cref{sec:eval:telescope}), \sysname exports erroneous packets (possibly anonymized/truncated according to policy) with their timestamp.

Host liveness is tracked at the SDN controller using a per-host bitmap indexed by internal IP address (live: 1, dark: 0). The controller marks a host as \emph{alive} in the bitmap whenever it installs a long-lived MAT rule or short-lived BF rule for one of that host's flows, and it resets the bitmap every minute so that hosts that stop generating traffic transition back to \emph{dark}. The controller exports incremental bitmap updates to the backend collector. This enables offline analytics to distinguish scans targeting dark hosts from inactive ports on live hosts.
\end{review}

\section{\sysname in action}
\label{sec:action}

\smartparagraph{Setup}
We implemented \sysname with $\text{P4}_{16}$ for the switch, and C/C++/Python for the \acrshort*{nf} and controller.
We deploy \sysname on the SUP4RNET~\cite{sup4rnet} testbed, on a Debian 12 server equipped with 2 Intel Xeon Gold 6252N 24-core CPUs and 128 GB of RAM, connected to the P4 switch equipped with 6.5 Tbps Wedge 100BF-32X Intel Tofino ASIC~\cite{IntelTofinoSeries}.
We run \acrshort*{nf} and the SDN controller on the same virtual machine (VM) featuring 16 logical cores and 16 GB vRAM. 
The SDN controller leverages the \gls{bri} to program the Tofino switch and communicates via a gRPC~\cite{grpc} channel.
Other parameters, such as the DT timeout and BF rotation interval, are discussed in \cref{sec:perf}.
According to the Tofino compiler report~\cite{tna}, the P4 program of \sysname consumes 26.8\% SRAM, 42.0\% Map RAM
, and 0.7\% TCAM resources.

\smartparagraph{Campus network deployment}
We have deployed \sysname in a campus network for more than a year. The network is connected to the Internet with a 20~Gbps bidirectional link. At peak time, the border router forwards around 16~Gbps of total traffic. \sysname has collected data involving 34,304 internal IPv4 addresses.\footnote{The campus uses one /17 and six /24 subnets, summing 34,304 addresses.}
Most of these addresses are allocated to desktops and laptops that go intermittently offline. 
\begin{review}
\gls{nat}
\end{review} is heavily used for both the WiFi and some department networks.
Since \sysname observes traffic after NAT, outbound analysis reflects NAT-translated source IP addresses rather than individual hosts, while inbound erroneous traffic remains unaffected as scanners target publicly routable addresses.
Several campus servers offer regular services on the Internet. A border firewall protects all internal hosts, allowing incoming connections only to a subset of servers and services. As such, most internal hosts and services look unreachable from remote hosts. Additionally, a /23 subnet serves as a traditional telescope. It has been operating for more than 8 years to observe only Internet radiation.

\smartparagraph{Evaluation metrics and use cases} From the collected traffic traces, we evaluate the following metrics: 
    (i) the number of erroneous packets received;
    (ii) the number of unique external senders;
    (iii) the number of unique internal addresses contacted;
    (iv) the diversity of source ports used and destination ports targeted.
We compare these metrics between \sysname and a /23 dark address space (pure telescope), to quantify the additional visibility gained beyond traditional telescope-based monitoring. Unless otherwise stated, we aggregate statistics over 1-hour windows.
Beyond aggregate metrics, we show how these traces translate into actionable operational insights through representative incidents \sysname has observed in the wild, including
amplification attacks, external routing anomalies, service misconfigurations (see \Cref{sec:eval:telescope}), and compromised internal hosts, stale DNS configurations (see \Cref{sec:eval:internal_radiation}).

\smartparagraph{Traffic filtering} In our deployment, \sysname's filtering mechanism effectively suppresses over 90\% of the traffic, with the \acrshort*{nf} managing just 10\% of the total traffic.

\subsection{External Erroneous Radiation}
\label{sec:eval:telescope}

Unlike a pure telescope, \sysname monitors erroneous traffic generated by both \emph{dark} and \emph{live} hosts.  Here, we investigate whether and how the addresses monitored by \sysname attract different traffic compared to the pure telescope. 
\begin{review}
We leverage the host liveness metadata to distinguish between traffic received by live and dark hosts, as discussed in \cref{sec:impl:metadata}. 
\end{review}
For this study, we sample 25 days of data collected from December 7 to 31, 2024.

\begin{table}[t]
\centering
\resizebox{\columnwidth}{!}{%
    \begin{tabular}{c c c c c c c c }
    \toprule
    \textbf{Probes} & \textbf{Dst. IPs} & \textbf{Dst. Ports} & \textbf{Src. ASs} & \textbf{Src. IPs} & \makecell{\textbf{Acknowledged} \\ \textbf{Scanners}}   \\
    \midrule

    /23 telescope & 510 & 1k & 715 & 7k & 708 \\
    \textbf{\sysname} & 34304 & 62k &  2310 & 17k & 935 \\
    \bottomrule
    \end{tabular}
    }
\caption{Quantitative comparison between traffic collected by \sysname and pure telescope every hour.}
\label{tab:sender_quantiy}
\end{table}

\begin{figure}[t]
    \centering
    \includegraphics[width=0.9\columnwidth]{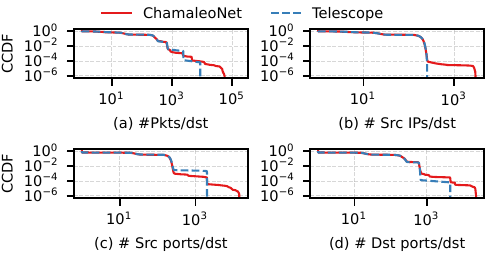}
    \caption{Distribution of Traffic Statistics (CCDF, Log–Log Scale). (a), (b), (c), and (d) denote the number of packets, unique source IPs, unique source ports, and unique destination ports observed per destination IP address, respectively.}
    \label{fig:src_ip_ccdf}
\end{figure}

\smartparagraph{Sender diversity and coverage}
In general, \sysname observes on average 17215 unique external senders and $1.21\times10^{7}$ erroneous flows.
\Cref{tab:sender_quantiy} shows the average amount of external erroneous traffic observed every hour. \sysname captures traffic from substantially more senders: $3.2\times$ more unique Autonomous System (AS), $2.4\times$ more unique IP addresses, and $1.3\times$ more acknowledged scanners.\footnote{Ground truth: \url{https://gitlab.com/mcollins_at_isi/acknowledged_scanners}} This sub-linear growth is expected because many large-scale scanning campaigns probe broad portions of the IPv4 space, making most scanners already visible in the /23 deployment. Expanding coverage therefore primarily increases the interaction intensity per scanner, while newly observed scanners are mainly from small, localized probing activities (see~\Cref{fig:src_ip_ccdf} in the following). Yet, senders in \sysname are also significantly more aggressive, contacting almost the whole port range ($62\times$).

\smartparagraph{Sender aggressiveness}
We analyse the distribution of traffic statistics in~\Cref{fig:src_ip_ccdf}, including (a) packets, (b) distinct source IPs, (c) distinct source ports, and (d) distinct destination ports. To ensure a fair comparison between \sysname (/17) and the /23 telescope, all metrics are computed on a per-destination-IP basis.

Given the fact the most of addresses monitored by \sysname is \emph{dark} in practice, we expect their traffic statistics to be similar to the pure telescope, as reflected by the overlapping CCDF curves. However, \sysname exhibits consistently heavier tails across all four metrics: the top $0.1\%$ of destinations receive substantially more packets (up to $14\times$), from more distinct senders (up to $11\times$), with greater source-port diversity (up to $8\times$) and broader destination-port coverage (up to $3\times$). \Cref{fig:src_ip_ccdf}(b) reveals stronger concentration effects: certain hosts in \sysname are contacted by significantly more unique senders. \Cref{fig:src_ip_ccdf}(c) indicates more aggressive probing behaviours, consistent with scanners randomizing source ports or issuing repeated connection attempts. \Cref{fig:src_ip_ccdf}(d) suggests that some destinations are subjected to wider port coverage, attributing to deeper or more comprehensive scanning activity. Overall, erroneous traffic collected by \sysname exhibits consistently heavier-tailed per-destination distributions, showing increased scanner aggressiveness in terms of traffic volume and probing breadth.

\begin{figure}[t]
    \centering
    \includegraphics[width=\columnwidth]{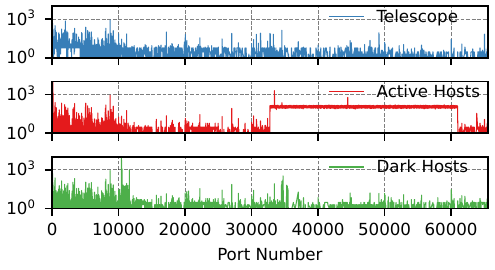}
    \caption{Erroneous packets per destination port on the telescope, active hosts and dark hosts.}
     \label{fig:practical_deployment_ports}
\end{figure}

\cref{fig:practical_deployment_ports} shows the number of packets per port observed in the top three Telescope receivers (top), active receivers (middle), and dark receivers (bottom) within \sysname over one day. The figure clearly shows that external senders target different port ranges with different intensities. While well-known ports and low port numbers receive more radiation, the presence of active services causes a significant increase in unsolicited traffic, causing the per-port probability to change. 

In a nutshell, \sysname collects much more Internet radiation than a pure telescope. As~\cite{PauleyDScopeCloudNativeInternet2023,RichterScanningScannersSensing2019a} transformed a CDN and a cloud provider into a telescope,  \sysname transforms any live network into a better telescope, without the need to reserve precious IP addresses for this.

\begin{figure}[!t]
    \centering
    \includegraphics[width=\columnwidth]{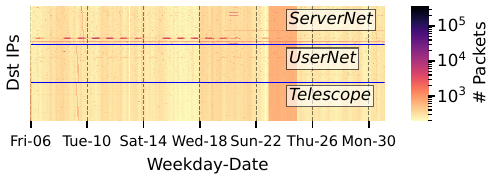}
    \caption{Incoming traffic collected by \sysname and pure telescope (1-hour interval). Blue lines mark the three /24 subnets.}
    \label{fig:temporal_traffic}
\end{figure}

\smartparagraph{Temporal evolution}
We now examine the temporal evolution of external erroneous radiation using two representative /24 subnets monitored by \sysname. \textit{ServerNet} is a /24 subnet hosting 10-15 campus servers, the other addresses acting as dark hosts. \textit{UserNet} is a /24 subnet with client hosts that are unreachable from outside. For comparison, we also include one /24 telescope subnet. \Cref{fig:temporal_traffic} illustrates the erroneous packets for each address in \textit{ServerNet}, \textit{UserNet}, and \textit{Telescope}, separated by solid blue lines. 

Scan pattern is very similar, with different intensities -- notice the 48-hour intensive scanning event during Dec.~23--24 period. Yet, the \textit{ServerNet} shows clearly some different pattern: the continuous horizontal line refers to active servers that external actors keep scanning on several ports.
Some intermittent pattern related to human activity emerges. Here, thousands of external addresses send erroneous packets to the public addresses allocated to the campus WiFi network NAT. These packets are due to ongoing connections where the internal WiFi host suddenly becomes unreachable, and the external senders keep retransmitting packets. 
Additionally, in the \textit{ServerNet} we observe some sudden and prolonged increase in erroneous traffic (small red horizontal bars). These are external scanners that look for other servers in this /24 subnet.

\smartparagraph{External senders targeting campus addresses}
During the study period, we observed a total of 436k unique external senders. Among them, 56.7\% target campus addresses exclusively and avoid any telescope address (while only 0.3\% interact solely with the telescope). By analysing the top-senders, we identify a variety of behaviours, including attack patterns, scanning activities, misconfigurations, etc. We describe some notable findings below (with anonymized addresses).

\begin{itemize}[noitemsep, topsep=0pt,leftmargin=*]
    \item IP 5.96.X.X: This is the most active sender. It generates an average of 3,000 packets per hour for the entire period. All packets go to UDP port 123 on a specific internal server, strongly suggesting a Network Time Protocol (NTP) attack. 
    
    \item IP 188.92.X.X: It scans ports 37215, 52869, and 49152 on thousands of hosts, hinting at a Satori bot~\cite{al2022nospring}. It also scans other ports (e.g., 1900 and 2048) indicating broader scanning. The sender is present in blocklists.
    
    \item IP 193.X.X.X: It belongs to the campus network provider. It keeps sending more than 1,000 ICMP Time Exceeded messages per hour, targeting 336 internal addresses. This suggests a routing issue (or blackholed destination). Internal clients cannot reach the final destination.
    
    \item IP 94.143.X.X: It consistently contacts a port on an internal server from source port 57498, sending about 720 packets per hour. This indicates a misconfigured or failed service.
    
    \item 38 IP addresses (e.g., 49.232.X.X) conduct  synchronised ICMP scans throughout the period, sending an average of 394 ICMP Echo Requests per hour to 20 internal IP addresses. The campus firewall blocks ICMP echo requests.
\end{itemize}

\begin{figure}[!tb]
    \centering
    \includegraphics[width=\columnwidth]{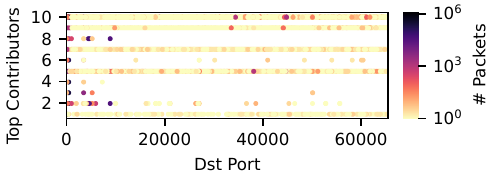}
    \caption{TCP destination ports of internal erroneous traffic generated by the top-10 contributors.}
    \label{fig:internal_radiation}
\end{figure}

\smartparagraph{Firing impersonators to engage with scanners}
To understand which services the scanners are interested in, we activate TCP impersonator on some unused IP addresses on the campus for three hours. These impersonators complete the TCP three-way handshake, observe the first application message, if any, and tear down the connection.
For impersonation traffic, \sysname operates in a pass-through mode with explicit redirection. The switch bypasses filtering and anonymization and forwards traffic to \acrshort*{nf} directly. In this mode, \acrshort*{nf} disables buffering and detection logic and forwards traffic to the responders, while simultaneously mirroring a copy to the collector for deeper analysis. 
We run nDPI~\cite{deri2014ndpi} to extract the application protocol classification of captured traffic, focusing on traffic to ports in the $[30000, 61000]$ range. Results show a mix of different protocols, the majority unknown to nDPI (45\%) (because only the first payload is captured), or has no payload (21\%) (because of client-initiated protocol), with HTTP (18\%), TLS (4\%), LDAP, DRDA, DRP, SQL, PRTP and other protocols being probed ($<$1\%). This confirms that the external scanners probe for various services on non-standard ports once they find the host alive~\cite{SoroEnlighteningDarknetsAugmenting2023}. The impersonating ability of \sysname is fundamental for exploring the intention of senders.


\subsection{Internal Erroneous Radiation}
\label{sec:eval:internal_radiation}

Let us focus on the erroneous packets generated by internal hosts that \sysname captures. Recall that a pure telescope will not collect any information on this.
On average, we observe approximately 40\,k outgoing erroneous packets per hour, consisting of roughly 40\% TCP packets, 36\% UDP packets, and 24\% ICMP packets. Again, we use simple analytics to quickly highlight suspicious behaviours.

\smartparagraph{Top senders of internal erroneous traffic}
We look at those hosts that contribute the most to the erroneous radiation. We consider one day of traffic. Each of the top senders contributes tens of thousands of erroneous packets. \cref{fig:internal_radiation} shows the destination ports targeted by the top-10 senders, the most active on the bottom.
Two patterns emerge: (i) horizontal scan patterns that target all (e.g., 1st, 4th-6th senders) or most ports (e.g., 9th sender); (ii) vertical scan patterns, targeting few ports (e.g., 3rd, 7th senders).
Some vertical scanners constantly send thousands of unanswered requests per hour to very few IP addresses in a cloud system. This hints at misconfigured systems that keep sending traffic to non-existent servers (some examples later).
For horizontal scanners, some target a few external hosts while others contact hundreds of different destinations. This hints at bots or malicious scanning activities that we signalled to our IT security team.


\begin{figure}[t]
    \centering
    \includegraphics[width=\columnwidth]{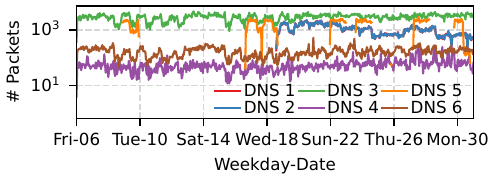}
    \caption{Erroneous DNS traffic sent by the campus DNS servers to external DNS servers.}
    \label{fig:dns}
\end{figure}
\smartparagraph{Erroneous DNS traffic}
By looking at the most targeted ports, we notice a lot of erroneous UDP traffic destined to port 53/DNS and sent by the campus official DNS resolvers. Digging into this, we observe that most of this traffic is directed to a few authoritative DNS resolvers (in Iran, China, India), which never responded to our resolvers. We report the temporal evolution of this erroneous traffic in Fig.~\ref{fig:dns}. We suspect this suspicious activity is part of attacks where infected internal clients send DNS requests to our resolvers that route them to the authoritative resolvers. Again, we signalled the incident to our IT security team.

\smartparagraph{Other findings}
Among the recurrent patterns exposed by \sysname, we observe occasional peaks of unanswered traffic to popular external servers. Unlike the previous cases, these patterns stem from benign communications but exhibit erroneous behaviour, likely due to misconfigurations or transient operational issues. Since \sysname targets operational visibility rather than attack detection, we treat these events as visibility signals of abnormal network conditions.
For example, we observe a steady stream of TCP SYN packets, about 3.7k per hour, from a specific internal server to a remote HPC repository. Investigation revealed that the repository had been decommissioned, while its DNS name remained registered, causing the server to keep reconnecting.

Another case involves over 50 internal hosts generating unusual ICMP port unreachable messages. We traced them to DNS acceleration tools that send parallel queries to multiple open resolvers and close the socket after the first response. Later responses then arrive at closed sockets, causing ICMP port unreachable messages that \sysname correctly logs as erroneous traffic.

Understanding the precise reasons for these behaviours is outside our scope. However, these simple findings illustrate the benefits of the visibility \sysname provides.


\section{System Performance Evaluation}
\label{sec:perf}

We now present results collected in controlled experiments to optimize \sysname design, select system parameters and characterize scalability.

\subsection{Parameter Setting} 
\label{sec:perf:param}

We set up a second VM to run a traffic generator, bridged to the server E810 NIC in PCIe passthrough mode using a separate SR-IOV~\cite{sr-iov} virtual function. Packets transmitted from the traffic generator VM are forwarded to the P4 switch before returning to the \acrshort*{nf} VM.

\smartparagraph{Replay traffic workload (TW-R)} 
We run a traffic replay workload that uses realistic traffic patterns captured from a campus network, containing about 6 million flows and 152 million packets over 10 minutes. We use up to 10 \texttt{tcpreplay}~\cite{tcpreplay} instances to generate different traffic rates, splitting and multiplexing traces in parallel to mimic real traffic patterns. \begin{review}We label this workload as \textsf{\small TW-R} for later reference.\end{review}

\begin{figure}[t]
    \centering

    \includegraphics[width=0.6\columnwidth]{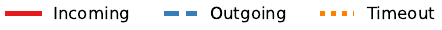}

    \begin{minipage}[t]{0.319\columnwidth}
        \centering
        \includegraphics[width=\linewidth]{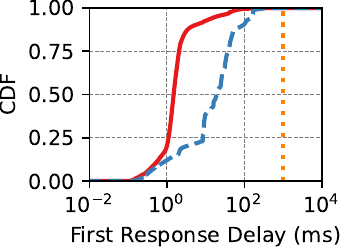}
        \subcaption{TCP}
        \label{fig:tcp_response_delay}
    \end{minipage}
    \begin{minipage}[t]{0.319\columnwidth}
        \centering
        \includegraphics[width=\linewidth]{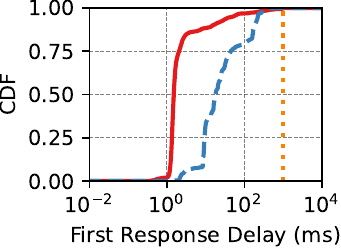}
        \subcaption{UDP}
        \label{fig:udp_response_delay}
    \end{minipage}
    \begin{minipage}[t]{0.319\columnwidth}
        \centering
        \includegraphics[width=\linewidth]{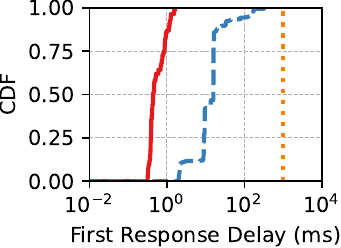}
        \subcaption{ICMP}
        \label{fig:icmp_response_delay}
    \end{minipage}

    \caption{First response delay of incoming and outgoing flows, measured in our campus network.}
    \label{fig:response_delay}
\end{figure}





\smartparagraph{DeTection (DT) timeout}\label{sec:detection-mechanism} For setting the DT timeout we observe the empirical response delay statistics of actual flows. 
We analyse the campus traces to measure the \emph{first} response delay of answered flow, i.e., the time elapsed between the first packet and the first response packet. We observe more than 100,000, 72,000 and 1,000 TCP, UDP and ICMP flows, respectively. About one-tenth are incoming, i.e., requests launched by external clients targeting internal services. 

Fig.~\ref{fig:response_delay} shows the Cumulative Distribution Function (CDF) of the response delay for TCP, UDP, and ICMP, respectively. The CDFs show that about 80\% of incoming requests are answered in a few ms. This is expected since internal hosts located within the main campus LAN have a short Round Trip Time (RTT). Because our campus also includes some remote laboratories connected with \begin{review}
Virtual Private Networks (VPNs)
\end{review} over the Internet, some clients and servers suffer longer RTT.
Looking instead at outgoing flows, the response time is significantly higher as it includes the RTT to servers located anywhere on the Internet. 

The choice of a proper DT timeout is a balance between accuracy and system load: increasing the timeout improves flow identification accuracy, but impacts system load and scalability.
In our setting, we set the timeout to 1s to correctly classify over 99.9\% incoming and 99.8\% outgoing answered flows. Note that \sysname could be easily extended to support multiple timeouts by simply splitting traffic to separate NFs or packet descriptor queues, e.g., using different DTs for incoming and outgoing requests, as done in telescope-like systems~\cite{GriffioenHaveYouSYNMe2024, HiesgenSpokiUnveilingNew2022, RichterScanningScannersSensing2019a}.

\begin{figure}[!t]
    \includegraphics[width=\columnwidth]{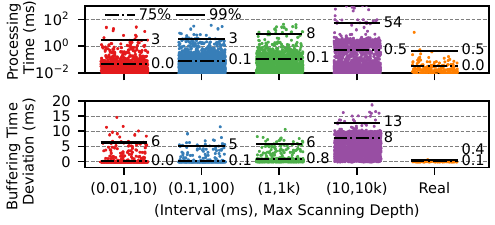}
    \caption{Periodic check of timeout expiration for various configurations of $(P_{DT}, d_{\max})$, with DT timeout set to $1s$. ``Real'' refers to the deployment in our campus network.} 
    \label{fig:periodic_check_performance}
\end{figure}

\begin{review}
\smartparagraph{DT timeout expiration routine: tuning ($P_{DT}$, $d_{\text{max}}$)}
As elaborated in Sec.~\ref{sec:impl}, \pktrx handles both incoming suspicious packets and DT timeout expiration. Every $P_{DT}$ seconds, packet reception is briefly interrupted to check the HoL descriptor and pop expired entries from the ring queue, up to a maximum scanning depth $d_{\max}$. Longer checks increase the time packets spend waiting in the \texttt{libpcap} buffer, although this does not affect correctness as long as no packet is dropped.

We evaluate the impact of $(P_{DT}, d_{\max})$ on per-packet processing time and DT buffering accuracy, shown in the top and bottom subfigures of \Cref{fig:periodic_check_performance}, respectively. We sample one out of every $10,000$ packets and, under 10x \texttt{\small tcpreplay} sender load, configure all settings to perform timeout expiration checks at the same aggregate rate, i.e., $d_{\max}/P_{DT}=1$~M~checks/s. \Cref{fig:periodic_check_performance} reports the 75th and 99th percentiles across different configurations.

Frequent, fine-grained checks provide the best tradeoff. With $P_{DT}=0.01$~ms and $d_{\max}=10$, 99\% of packets are processed within 3~ms with no packet loss, compared to 55~ms when checking up to 10,000 descriptors every 10~ms. The buffering-time results show negligible deviation from the predefined DT timeout: for example, $P_{DT}=0.1$~ms and $d_{\max}=100$ keeps 99\% of packets buffered for about 1.005~s. Results with real traffic traces confirm the same trend.

These results reveal a broad stable operating region for $(P_{DT}, d_{\max})$. We therefore adopt fixed parameters selected empirically from this region.
\end{review}

\begin{review}
\subsection{Scalability Analysis}
\label{sec:eval:nf-tput}

In this section, we evaluate \sysname's performance under adversarial conditions that stress both the packet processing and rule management logic.

\smartparagraph{Adversarial traffic workloads}
We create two synthetic adversarial traffic workloads generated with Cisco T-Rex~\cite{TRex}, a state-of-the-art traffic generator in DPDK~\cite{dpdk}:\end{review}

\begin{itemize}[noitemsep, topsep=0pt,leftmargin=*]
\item \emph{DoS traffic} \begin{review}(\textsf{\small TW-DoS}):\end{review} every packet corresponds to a new flow with no response. This workload stresses \acrshort*{nf}'s packet processing throughput by maximizing the number of concurrent erroneous flows.
\item \emph{High-churn transient benign traffic} \begin{review}(\textsf{\small TW-HCB}):\end{review} every flow consists of exactly two bidirectional packets, a legitimate request and its corresponding response. This workload stresses both the \acrshort*{nf}'s packet processing throughput and the rule management logic on the fast-path, by rapidly generating new, short-lived benign flows. 
\end{itemize}
\begin{review}
For both workloads, we generate traffic with minimum packet size to maximize the packet processing load on \acrshort*{nf} (i.e., 64 bytes on Ethernet).
\end{review}

\begin{figure}[t]
    \centering
    \includegraphics[width=0.9\columnwidth]{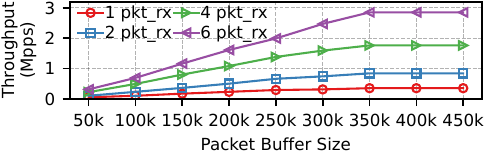}
    \caption{
    \begin{review}
        Throughput of the \acrshort*{nf} under DoS workload.
    \end{review}
    }
    \label{fig:throughput_dos}
\end{figure}

\begin{review}
\smartparagraph{\acrshort*{nf} packet processing throughput}
The first threat to scalability is \acrshort*{nf}'s packet processing throughput. When all incoming packets correspond to new flows, \acrshort*{nf} receives large bursts of packets, and high memory pressure to store flow states.
We benchmark \acrshort*{nf}'s peak throughput following RFC~2544~\cite{rfc2544}, stopping at the first packet loss. 
\Cref{fig:throughput_dos} shows the peak throughput achieved with 1, 2, 4, and 6 \pktrx threads, under different sizes of the ring buffer in the \acrshort*{nf}. As the DoS and high-churn transient benign workloads show similar results, we omit the latter for brevity.
For small packet buffer sizes, the throughput is limited by early buffer fill-up. 
For buffer size larger than $\sim$350k packets, the overheads for packet processing operations become dominant (e.g., buffering, expiry checking, and garbage collection). Packets cannot be drained from the ring buffer fast enough, resulting in packet drops and throughput saturation.
Overall, \acrshort*{nf} reaches a peak throughput of 350, 840, 1760, and 2850~kpps with one, two, four, and six threads, respectively, consuming about 16 GB of vRAM with six threads. 
This is 519$\times$ the erroneous traffic we normally receive in our deployment at the NF.
These results demonstrate linear scaling of \sysname's throughput when packet processing is parallelized across six \pktrx threads. 
\end{review}

\begin{figure}[t]
    \centering
    \begin{subfigure}[t]{0.49\columnwidth}
        \centering
        \includegraphics[width=\linewidth]{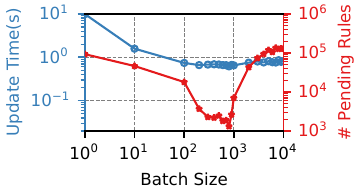}
        \subcaption{}
        \label{fig:pending-rules}
    \end{subfigure}
    \hfill
    \begin{subfigure}[t]{0.49\columnwidth}
        \centering
        \includegraphics[width=\linewidth]{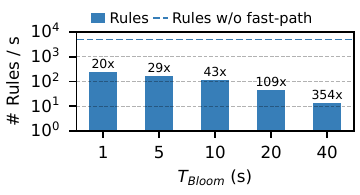}
        \subcaption{}
        \label{fig:slow_path_rate}
    \end{subfigure}
    \caption{
    \begin{review}
        (\subref{fig:pending-rules}) Batching amortizes gRPC call overhead, until the switch's processing time for rule installation becomes dominant.
        (\subref{fig:slow_path_rate}) The fast-path cuts the update rate on the slow-path by orders of magnitude, mitigating bottleneck \emph{\#1}.
    \end{review}
    }
\end{figure}

\begin{review}


\smartparagraph{Filtering rule management}\label{sec:eval:rule-scalability}
The second threat to \sysname's scalability is represented by the rule management mechanism. 
Here, scalability is constrained by two bottlenecks: \emph{(i)} the control-plane channel bandwidth for rule updates and \emph{(ii)} the switch's memory capacity to store flow rules. We analyse each of these bottlenecks in turn, highlight when they emerge, and show that \sysname's design effectively mitigates both.

\emph{Bottleneck \#1: gRPC channel for slow-path rule updates.} Under traffic with a prevalence of long-lived flows, the  gRPC control-plane channel may become the bottleneck due to the high rate of rule updates.
\sysname's design includes a fast-path filtering mechanism to mitigate this issue.
We quantify its effectiveness as follows. 

First, we run a controlled micro-benchmark to measure the maximum throughput for rule update on the slow path (i.e., gRPC channel). We run a control-plane script that continuously installs rules on the Tofino ASIC via the BFrt~\cite{bfrt} control-plane APIs.  
To maximize throughput, we batch multiple rule updates in the same gRPC request. We empirically tune the batch size to a value that amortizes the cost of the gRPC calls (maximizing throughput), without incurring excessive latency from bulk rule installation at the switch. Based on the measurements shown in \cref{fig:pending-rules}, we set the batch size to 200 rules, a value chosen in the sweet spot region.
With this value, we measure a maximum throughput of around 14k rule updates/sec, which we take as a reference capacity of the gRPC channel for the following experiment.

Second, we run \sysname \emph{with} and \emph{without} the fast-path enabled. We generate a traffic workload that stresses the gRPC control-plane channel by generating a large number of long-lived flows, which require slow-path rule updates. To do so, we replay traffic \textsf{\small TW-R} and scale the number of replicas until we saturate the gRPC channel capacity, measured in the earlier experiment. 
When the fast-path is enabled, we set $n_{\text{longlived}}=2$ for the long-lived flow promotion (discussed in \cref{sec:bloomfilter:promotion}).
\cref{fig:slow_path_rate} shows
that the fast-path can mitigate the slow path bottleneck, by cutting the rule update rate by up to 354$\times$. 
Longer BF rotation intervals $T_{\text{bloom}}$ lead to lower slow-path update rates, as more transient flows are filtered in the fast path and do not require promotion to the slow path.
\end{review} 

\begin{review}
\emph{Bottleneck \#2: Switch memory usage.}
The second bottleneck is the switch's memory capacity to store flow rules. 
Hence, we quantify the impact of the fast-path scheme on the TCAM usage in \cref{fig:slow_path_usage_ts}.
The plot shows the number of rules installed in the switch's TCAM over time, with and without the fast-path enabled. 
With the fast-path enabled, the number of installed rules remains stable and low, peaking below 25k MAT entries (2.8\% of the 80k allocated entries), compared to 99.96\% without the fast-path.
This is because short-lasting flows that do not persist across multiple BF rotation epochs are not promoted, while only stable services are installed as MAT entries. 
We next examine whether moving transient benign-flow rules to the fast path suffers from a memory bottleneck. 
\end{review}

\subsection{Impact of Bloom Filter (BF) on Visibility}
\label{sec:filtering_accuracy}
\begin{figure}[t]
    \centering
    \begin{subfigure}[t]{0.49\columnwidth}
        \centering
        \includegraphics[width=\linewidth]{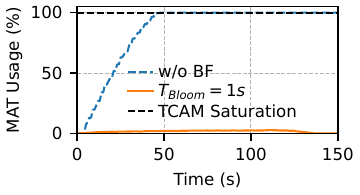}
        \subcaption{}
        \label{fig:slow_path_usage_ts}
    \end{subfigure}
    \hfill
    \begin{subfigure}[t]{0.49\columnwidth}
        \centering
        \includegraphics[width=\linewidth]{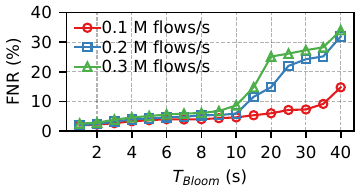}
        \subcaption{}
        \label{fig:empirical_fnr}
    \end{subfigure}
    \caption{
    \begin{review}
        (\subref{fig:slow_path_usage_ts}) The Bloom filter (BF) bounds TCAM occupancy, whereas the baseline saturates the capacity shortly.
        (\subref{fig:empirical_fnr}) Empirical false negative rate of \sysname’s filtering across different $T_{\text{bloom}}$ under adversarial traffic.
    \end{review}
    }
\end{figure}

\begin{review}
Storing benign flow rules installed via the fast-path on Bloom filters consumes SRAM resources. Because SRAM memory is scarce in programmable switches (a few dozens of MB on Tofino), here we evaluate the trade-off between memory usage and visibility of erroneous traffic.

Increasing $T_{\text{bloom}}$ keeps benign-flow rules in the BFs for longer and therefore increases BF occupancy. BF occupancy determines collision probability in the BF and drives the visibility trade-off: a false positive in any of the $r$ BFs can incorrectly match an erroneous packet to a benign flow entry, causing the packet to be discarded, i.e., a false negative for \sysname
\footnote{The false negative rate (FNR) in \sysname can be computed as $\text{FNR} = 1 - (1 - \text{FPR}_{BF})^{r}$, where $\text{FPR}_{BF} = (1 - e^{-kn/m})^k$ is the standard BF false positive rate for $k$ hash functions, $n$ elements, and $m$-bit array.}.




\smartparagraph{Empirical \gls{fnr}} 
\Cref{fig:empirical_fnr} evaluates the visibility degradation of \sysname under adversarial traffic by reporting the empirical false negative rate. 
We construct a traffic workload that mixes the replay traffic workload \textsf{\small TW-R} with different proportions of high-churn transient benign traffic \textsf{\small TW-HCB}. We then measure an empirical FNR that quantifies the amount of ``lost'' erroneous traffic, \emph{i.e.}, erroneous traffic mistakenly dropped by the BFs. To measure the empirical FNR, we first identify the ground-truth erroneous traffic in the workload we generate, then we compare it with the erroneous traffic logged by \sysname.
We use $r{=}2$ BFs, each with $2$ hash functions and a bit array of $2^{22}$ entries. We set these parameters to the largest values allowed by the compiler and per-stage resource constraints of Tofino.
As shown in \Cref{fig:empirical_fnr}, the FNR grows with $T_{\text{bloom}}$ and the proportion of high-churn benign traffic, confirming the expected trade-off. 
At a churn rate of 0.3~M flows/s, \emph{i.e.}, around the saturation point of a single \pktrx thread, setting $T_{\text{bloom}}=1s$ leads to FNR 2.4\%.
In practice, the actual flow churn is over an order of magnitude lower on average in a campus network (see operating point below).
In addition, we observe a region, for $T_{\text{bloom}} \le 10s$, where the empirical FNR grows slowly, and remains at acceptable levels. This suggests that \sysname can be configured to operate in a regime where the BF provides effective filtering while keeping visibility degradation under control. 
 
\medskip
\emph{\smartparagraph{Operating point (campus network)} In our real deployment, the benign-flow churn rate is around 11.8~k flows/s on average. We set $T_{\text{bloom}}=5s$, which corresponds to an empirical FNR of 5.0\% under the adversarial workload.  With this setting, the rule update rate on the slow-path is reduced by 36.1\%, resulting into 2454~rules/sec on average. Correspondingly, the load on the \acrshort*{nf} is reduced by 90.0\%, resulting into 76.4~kpps received traffic, which can be handled by one \pktrx thread.}
\end{review}




\section{Discussion}
\label{sec:discussion}

\begin{review}
\smartparagraph{Larger deployments} We evaluated a single switch deployment, and demonstrated that our current design suffices for a campus-network size.
    To scale to larger deployments, such as ISP or data center size, network managers could shard \sysname's dataplane across multiple P4 switch replicas, to further scale out SRAM memory for BFs, and gRPC channel capacity for long-rule installation. \sysname's design is compatible with sharding. Each switch replica can run the same P4 program, and flows can be distributed across replicas using standard affinity-preserving per-flow load balancing. The control-plane can also be replicated to manage each switch independently.
    Lastly, and orthogonally to this paper, systems optimizations~\cite{fajita} for high-speed stateful network functions can be adopted to further improve scalability of \acrshort*{nf} under multi-threading, by mitigating memory access contentions that inevitably arise as thread count increases.
    We regard these as implementation challenges rather than a fundamental scalability limit of \sysname's design.

\end{review}

\smartparagraph{Offloading \acrshort*{nf} to programmable dataplanes} 
Moving FSD-NF entirely into the data plane could eliminate the software FSD-NF bottleneck entirely. However, this design faces two fundamental challenges. 
First, simultaneously buffering suspicious packets of multiple flows for \emph{tens of milliseconds} is infeasible at modern link speeds, given the limited on-chip memories. For example, a 10MB buffer, compatible with a typical data center switch memory size~\cite{TEA,screamConext}, would fill in about 1ms at 100Gbps.
Second, programmable switch ASICs lack random-access memory, making it non-trivial to retrieve a buffered suspicious packet once it should be discarded. Integrating \sysname with state-of-the-art work~\cite{PayloadPark} is left for future enhancement.

\smartparagraph{In-path deployment} \sysname's architecture relies on mirroring ingress/egress traffic from the campus network to the SDN switch, resulting in an out-of-path deployment. Alternatively, \sysname could be deployed in-path on SDN-capable border router (or any downstream switch on the ingress/egress path). In such cases, the \acrshort*{nf} shall observe all suspicious packets and install a rule to \emph{forward} benign flow packets to the campus network, instead of \emph{dropping} them. 

In-path deployment enables the defense from erroneous flows, \emph{i.e.}, dropping or re-routing erroneous packets instead of forwarding them to the campus network. However, in-path deployment introduces extra challenges, such as invertible anonymisation and additional forwarding latency on regular traffic. It is also more vulnerable to false negatives: missing the detection of a benign flow would disrupt regular traffic, since no forwarding rule would be installed.
Consequently, network administrators may be reluctant to introduce \sysname on the live traffic path. The out-of-path design illustrated in \cref{fig:controller_arch} offers a less intrusive and pluggable solution. 


\section{Ethical Considerations}
\label{sec:ethical}

\smartparagraph{Privacy} Privacy is fundamental as legitimate traffic is also mirrored. We consider an ``honest but curious'' scenario where the person in charge of processing the data will not deviate from the defined goals. We thus adopt a \textit{data minimisation approach}: To minimise the information the controller and the collectors process and store, we keep only headers up to the transport protocol and remove any upper-layer information. We support IP address anonymisation for the internal network IP addresses by using a simple obfuscation mechanism. We leverage the data-plane programmability features to perform anonymisation directly at the switch (see Sec.~\ref{sec:anonymizer}).

\smartparagraph{Impersonating not-responding servers or services} \sysname cannot interfere with production traffic. Logging erroneous packets is not critical -- impersonating non-responding servers or services requires particular attention. When internal services go offline for maintenance or temporary failures, legitimate users could find themselves interacting with honeypots. Likewise, impersonating an internal offline client (e.g., a PC turned off at night) poses legal and security concerns if its address is converted into honeypot. We limit ourselves to demonstrating the \sysname impersonating ability with some pre-determined off-line addresses, leaving its extensive usage for future analysis. Notice that the privacy features must be disabled for the hosts/services \sysname will impersonate so that the original IP addresses and payload are exposed to the impersonator.

\section{Conclusions}
\label{sec:conclusion}

We presented \sysname, a system that transforms any production network into a programmable probe to monitor erroneous traffic -- such as unanswered, misrouted, or malformed packets. By leveraging SDN principles and programmable data planes, \sysname offers scalable, privacy-compliant traffic visibility without disrupting normal network operations. It operates transparently in live environments, collecting only erroneous traffic, anonymising sensitive information at the switch, and enabling both passive observation and active engagement through honeypot integration.

Our deployment over a year showed \sysname's effectiveness in uncovering external scanning behaviours and internal misconfigurations, outperforming traditional telescope setups in data richness and insights. It filtered up to 90\% of benign traffic at the hardware level, ensuring scalability with minimal resource requirements. \sysname is open source, modular, and ready for deployment in various environments, offering a practical and ethical solution to enhance network visibility and threat intelligence without requiring dedicated addresses or additional instrumentation.

\section*{Acknowledgment}
This work was supported by project SERICS (PE00000014) under the MUR National Recovery and Resilience Plan funded by the European Union -- NextGenerationEU and the PRIN ACRE (AI-Based Causality and Reasoning for Deceptive Assets -- 2022EP2L7H). This manuscript reflects only the authors' views and opinions and the Ministry cannot be considered responsible for them.

\bibliographystyle{IEEEtran}
\bibliography{bibliography}

\begin{IEEEbiography}[{\includegraphics[width=1in,height=1.25in,clip,keepaspectratio]{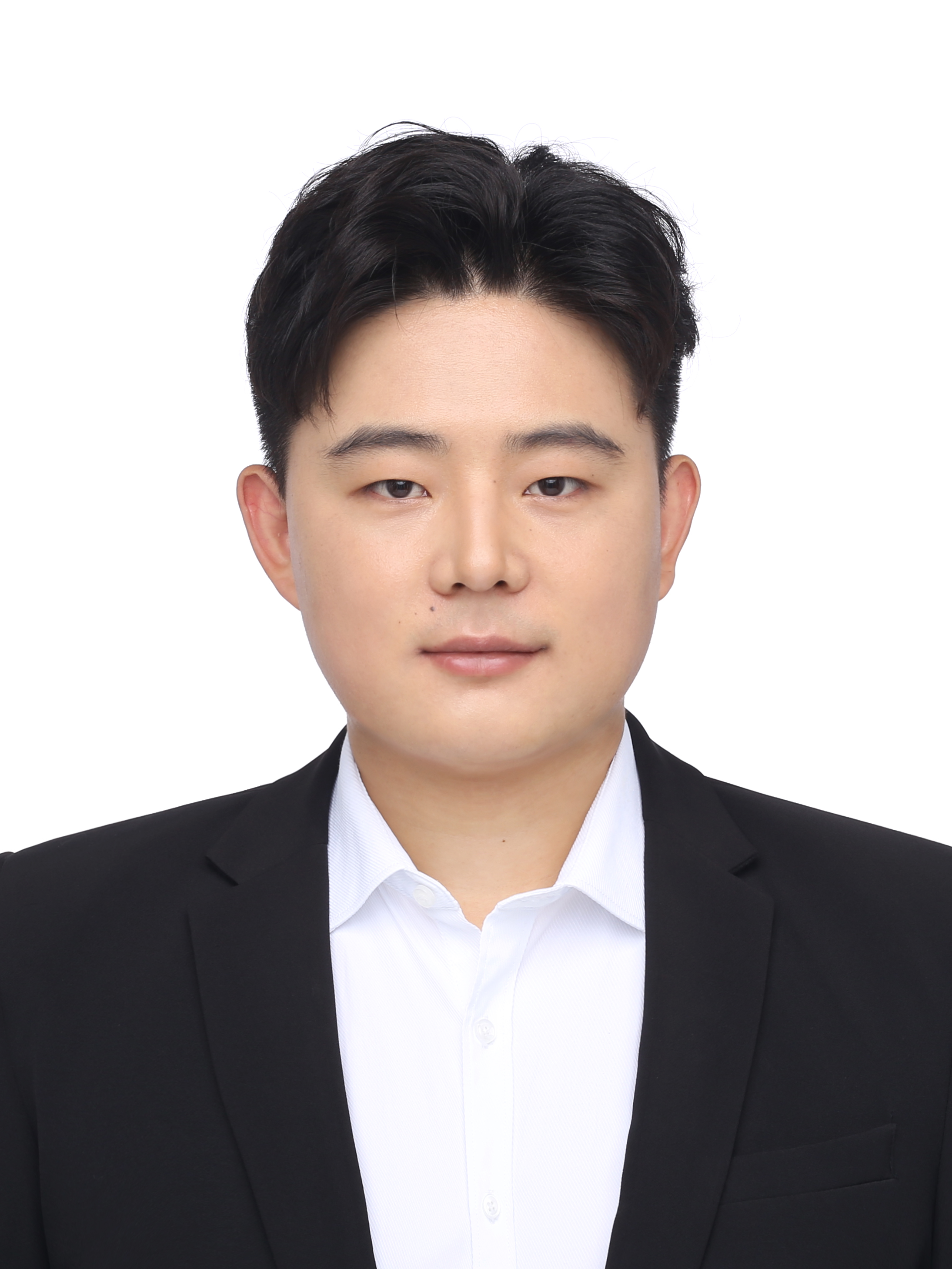}}]{Zhihao Wang}
is a Ph.D. student at the University of Electronic Science and Technology of China (UESTC), China. He visited Politecnico di Torino, Italy, funded by Chinese Scholarship Council (CSC). His research interests include network observability, programmable networks and cybersecurity.
\end{IEEEbiography}

\vskip -3\baselineskip plus -1fil
\begin{IEEEbiography}[{\includegraphics[width=1in,height=1.25in,clip,keepaspectratio]{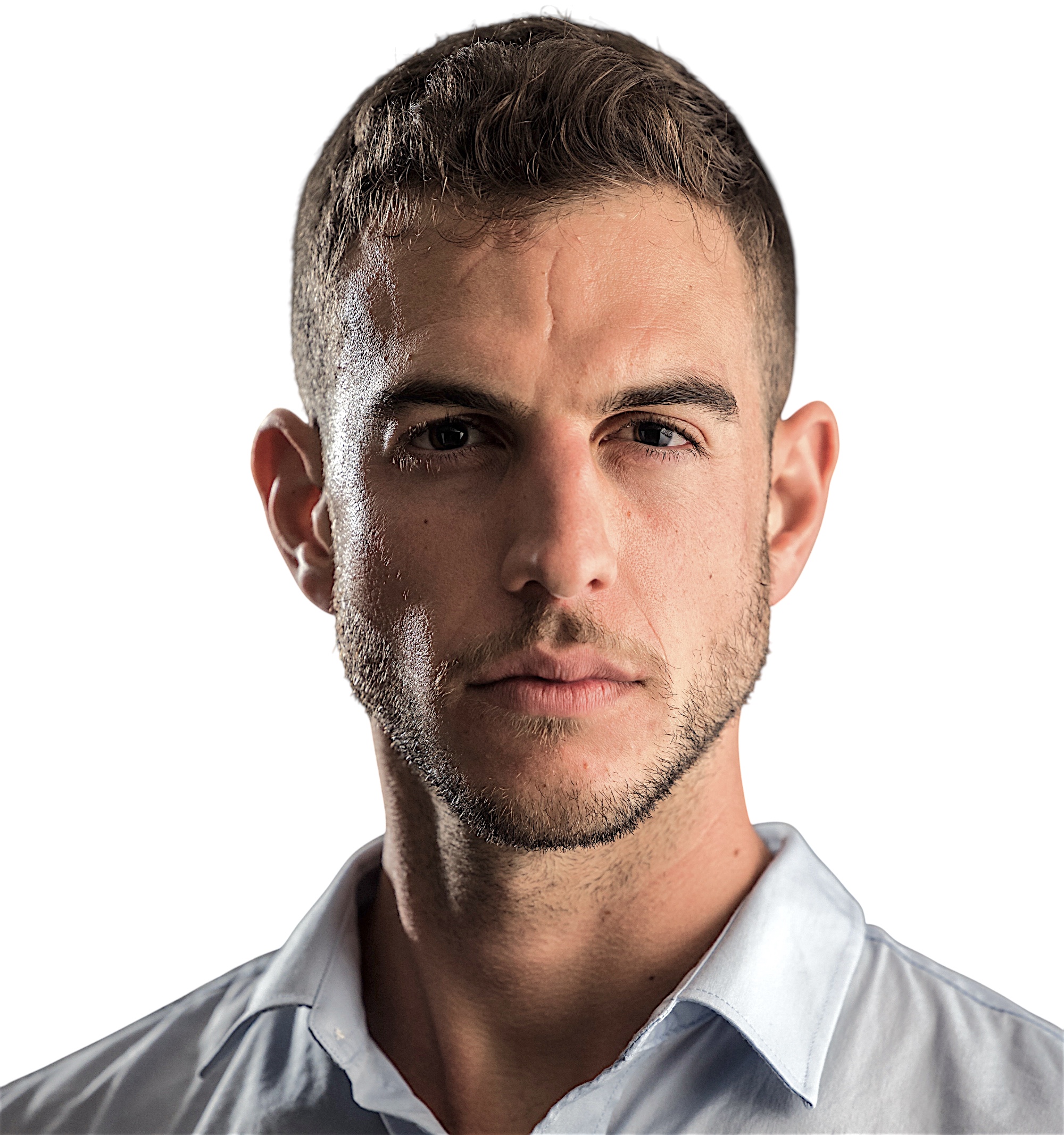}}]{Alessandro Cornacchia}
is a Postdoctoral Fellow at the King Abdullah University of Science and Technology (KAUST), Saudi Arabia. He obtained his Ph.D. from Politecnico di Torino, Italy.
His research interests span multiple aspects of distributed systems and data center networks, with a special appreciation for monitoring, observability and root-cause analysis. He served as a reviewer for ACM/IEEE journals and as a PC member for ACM workshops.
\end{IEEEbiography}

\vskip -3\baselineskip plus -1fil
\begin{IEEEbiography}[{\includegraphics[width=1in,height=1.25in,clip,keepaspectratio]{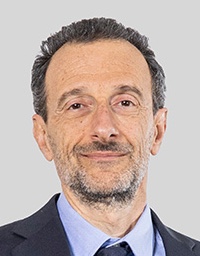}}]{Andrea Bianco}
(Senior Member, IEEE) received the Ph.D. degree in telecommunications from the Politecnico di Torino, Italy, in 1993. He is a Full Professor and the Vice Rector of the Internal Affairs, Politecnico di Torino. He has co-authored over 230 papers published in journals and presented at leading international conferences.
His research interests include design of all-optical networks, switch architectures for high-speed networks and data centers, SDN, and networked music performance.
\end{IEEEbiography}

\vskip -3\baselineskip plus -1fil
\begin{IEEEbiography}[{\includegraphics[width=1in,height=1.25in,clip,keepaspectratio]{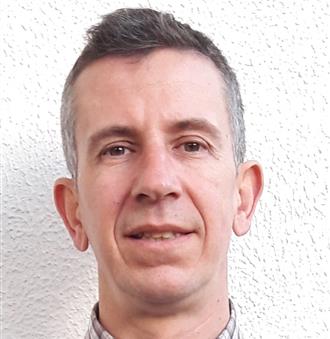}}]{Paolo Giaccone}
(Senior Member, IEEE) is a Full Professor with the Department of Electronics and Telecommunications, Politecnico di Torino, Italy.
During 2000--2001 and in 2002 he was with the Information Systems Networking Lab, Electrical Engineering Dept., Stanford University. His main area of interest is the design of network algorithms, in particular for SDN networks and cloud computing systems.
\end{IEEEbiography}

\vskip -3\baselineskip plus -1fil
\begin{IEEEbiography}[{\includegraphics[width=1in,height=1.25in,clip,keepaspectratio]{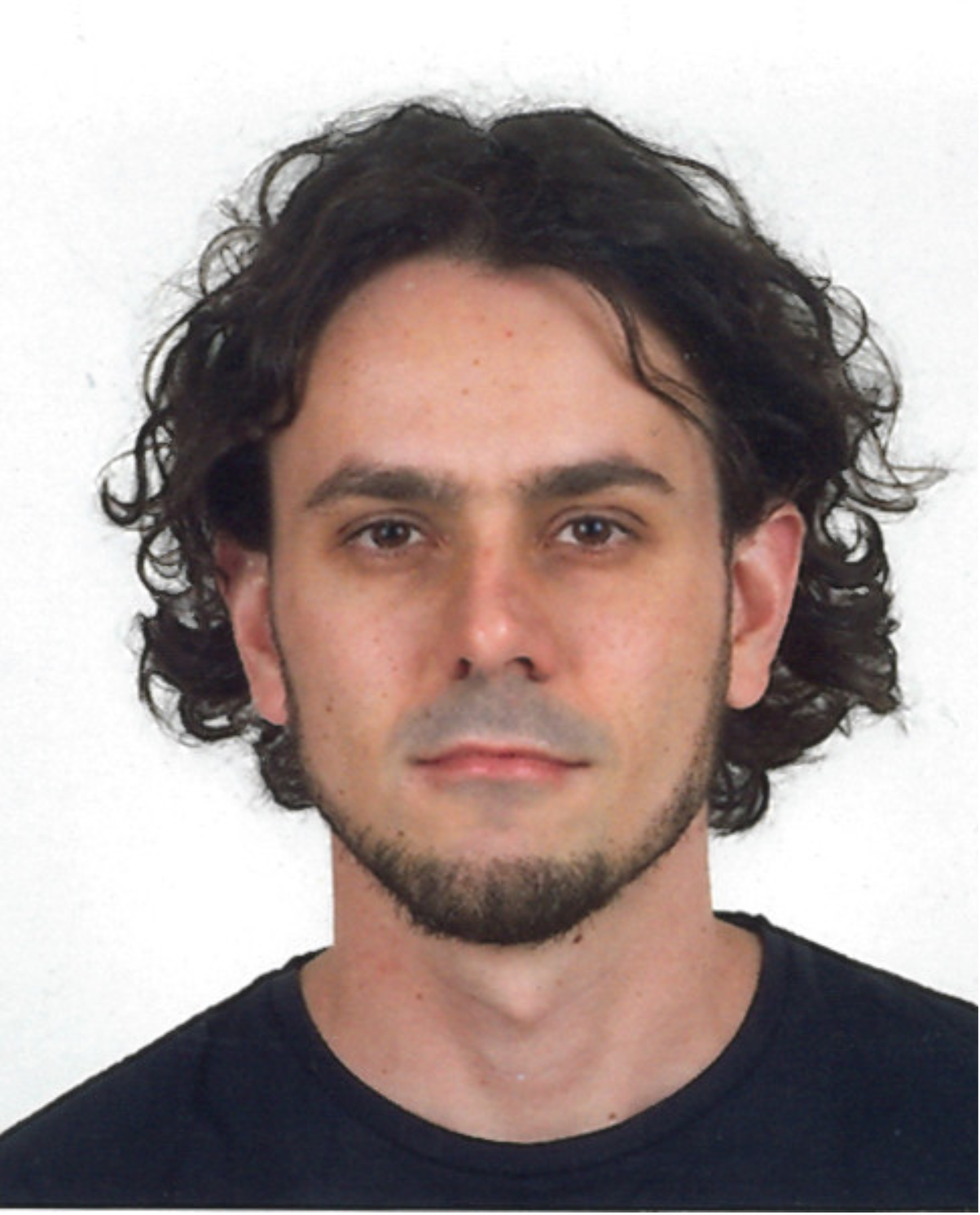}}]{Idilio Drago}
is an Associate Professor at the University of Turin, Italy. His research interests include network security, AI, machine learning, and Internet measurements. He is particularly interested in how AI and machine learning can help extract knowledge from network data, and secure the network.
Drago has a Ph.D. from the University of Twente, the Netherlands. He was awarded the IETF/IRTF Applied Networking Research Prize.
\end{IEEEbiography}

\vskip -3\baselineskip plus -1fil
\begin{IEEEbiography}[{\includegraphics[width=1in,height=1.25in,clip,keepaspectratio]{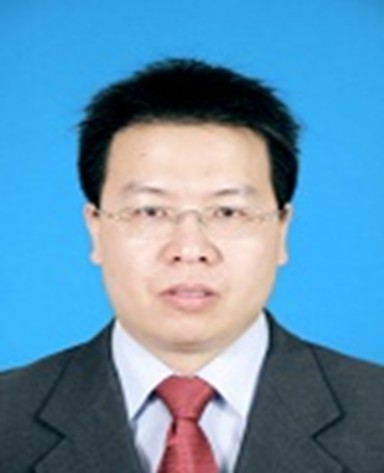}}]{Dingde Jiang} (Member, IEEE) is currently a professor with the School of Information and Communication Engineering, UESTC. His research focuses on network measurement, modelling and optimization, network management, network security, particularly in SDN, information-centric networking, energy-efficient networks, and cognitive networks. His research is supported by the NSFC, the Program for New Century Excellent Talents with the University of Ministry of Education of China, and so on.
\end{IEEEbiography}

\vskip -3\baselineskip plus -1fil
\begin{IEEEbiography}[{\includegraphics[width=1in,height=1.25in,clip,keepaspectratio]{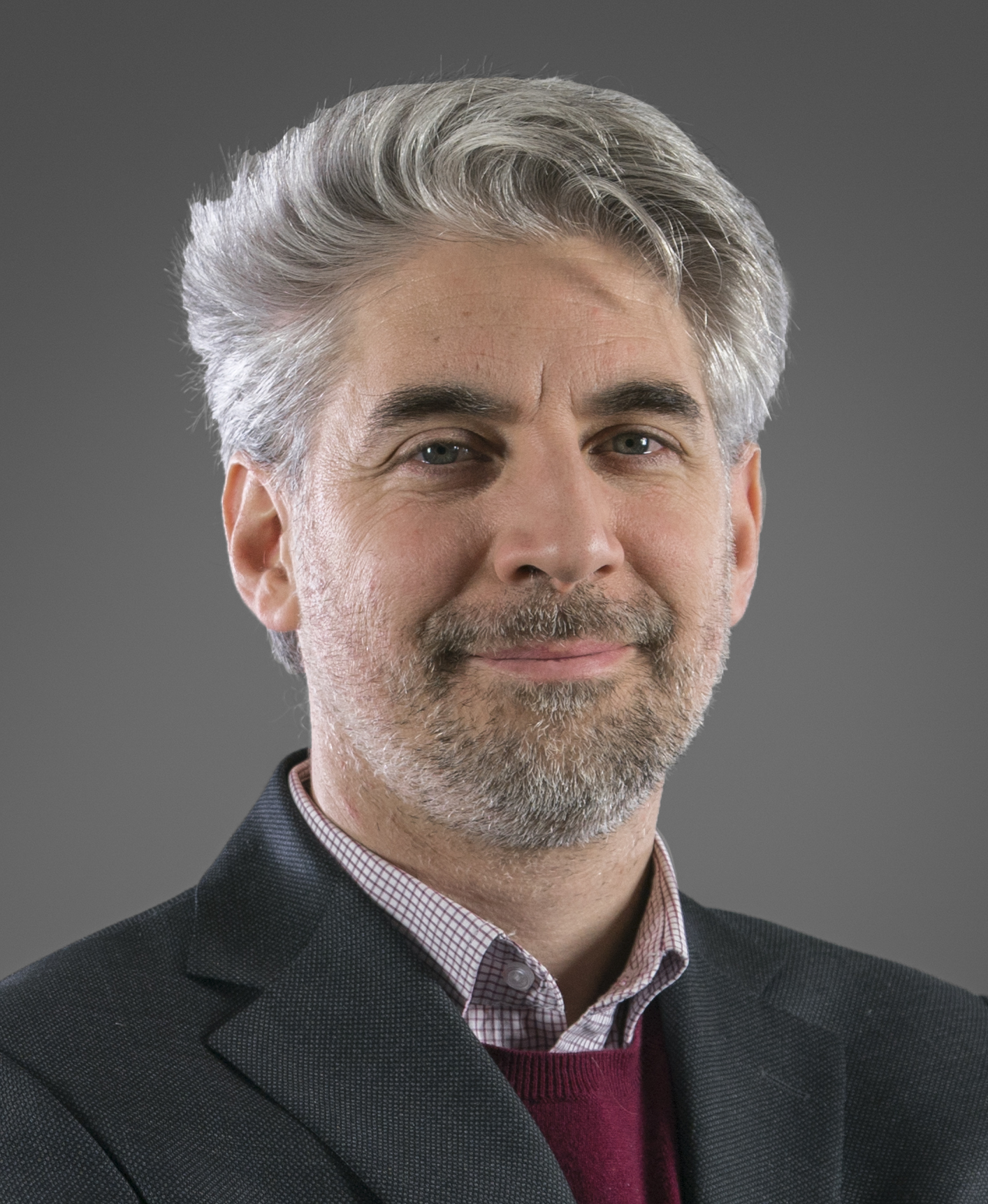}}]{Marco Mellia}
(F'21), Ph.D., is a full professor at Politecnico di Torino, Italy. He has co-authored over 300 papers published in journals and presented at leading conferences. He won the IRTF ANR Prize at IETF-88, and the best paper awards at IEEE P2P'12, ACM CoNEXT'13, IEEE ICDCS'15, ACM CCR'16, ITC'18. He is the Editor in Chief of the Proceedings of the ACM on Networking. His research interests are in the areas of Internet monitoring, users' characterisation, cyber security, and machine learning applied to different sectors.
\end{IEEEbiography}

\end{document}